\documentclass[a4paper,fleqn,usenatbib]{mnras}
\usepackage{epsfig}
\usepackage{amsmath}
\usepackage{natbib}

\newcommand{\be}{\begin{equation}}
\newcommand{\ee}{\end{equation}}
\newcommand{\ba}{\begin{eqnarray}}
\newcommand{\ea}{\end{eqnarray}}
\newcommand{\bi}{\begin{itemize}}
\newcommand{\ei}{\end{itemize}}
\newcommand{\bfi}{\begin{figure}
\epsfxsize=9cm
\epsffile}
\newcommand{\efi}{\end{figure}}
\newcommand{\no}{\nonumber}

\newcommand{\muk}{\mu {\rm K}}

\title[The ABS method]{{\bf ABS}: an {\bf A}nalytical method of {\bf B}lind {\bf S}eparation of CMB from foregrounds}
\author[Zhang et al.]{Pengjie Zhang$^{1,2,3,4}$\thanks{E-mail: zhangpj@sjtu.edu.cn}, Jun Zhang$^{1,4}$, Le Zhang$^{1,4}$\\
$^1$Department of Astronomy, School of Physics and Astronomy, Shanghai Jiao Tong University, 955
Jianchuan road, Shanghai, 200240\\
$^2$ IFSA Collaborative Innovation Center, Shanghai Jiao Tong
University, Shanghai 200240, China\\
$^3$ Tsung-Dao Lee Institute, Shanghai 200240, China\\
$^4$ Shanghai Key Laboratory for Particle Physics and Cosmology
}

\begin{document}
\maketitle
\begin{abstract}
Extracting CMB B-mode polarization from complicated foregrounds is a
challenging task in searching for inflationary gravitational
waves. We propose the ABS method as a blind and analytical solution to this
problem. It applies to the measured cross bandpower between
different frequency bands and obtains the CMB B-mode
bandpower analytically. It does not rely on assumptions of foregrounds
and does not require multiple parameter fitting. Testing against a variety of
foregrounds, survey frequency configurations and instrument noise, we verify its applicability
and numerical stability. The ABS method also applies to CMB
temperature, E-mode polarization, the thermal Sunyaev Zel'dovich
effect, spectral distortion, and even significantly different problems such as cosmic magnification. 
\end{abstract}
\section{Introduction}
Searching for  inflationary
gravitational waves \citep{Starobinskii79}  through the induced CMB B-mode polarization
\citep{1997PhRvL..78.2054S,1997ApJ...482....6S,1997PhRvL..78.2058K} is
a major endeavour of cosmology (e.g. BICEP:
\citet{BICEP2,BICEP3,2015PhRvL.114j1301B}; ACTpol: \citet{ACTpol};
SPTpol: \citet{SPTpol}; POLARBEAR: \citet{POLARBEAR2};  AliCPT:
\citet{AliCPT};  PIPER: \citet{PIPER};  CORE: \citet{CORE};  EPIC:
\citet{EPIC};  LiteBIRD:\citet{LiteBIRD}; PIXIE: \citet{PIXIE11}; PRISM:
\citet{PRISM};  PICO: \citet{2018SPIE10698E..46Y}).
It will open a window into the very
beginning of our universe. 

A major challenge of CMB B-mode detection is to accurately remove polarized galactic foregrounds
\citep{2015arXiv150201588P,2015arXiv150205956P,2015arXiv150606660P}.
At CMB frequency of $\sim 100$ GHz, a major
foreground  is the galactic thermal dust, which likely dominates
over CMB B-mode at $\nu\ga 100$ GHz, even for the cleanest sky areas
\citep{2015PhRvL.114j1301B}.  Synchrotron emission may be  another major
foreground, especially at lower frequency. Other polarized
foregrounds such as spinning dust
\citep{2011A&A...536A..20P,2015arXiv150201588P} and magnetic dust
\citep{2012ApJ...757..103D,2015A&A...576A.107P} may
also be non-negligible. 

Usually CMB  experiments rely on multi-frequency information to remove foregrounds
(e.g. \citet{2015arXiv150201588P,2015arXiv150205956P,2015arXiv150606660P,PIXIE11,PRISM}). This
kind of approaches faces a major uncertainty, namely the exact frequency
dependences of foregrounds and the exact number of independent
foreground components are unknown. For example,  recently Planck
found that dust foregrounds at  $217$ and $353$ GHz bands are
decorrelated at a few percent level, meaning the existence of multiple dust
components. This may lead to a significant bias in $r$ 
(tensor-to-scalar ratio)
\citep{2016MNRAS.458.2032R,2016arXiv160607335P,2016arXiv160608922P}. To
avoid such  potential bias, various methods blind of foregrounds, such
as the internal linear combination (ILC) method and the independent
component analysis (ICA) method, have been
constructed (e.g. the review article by
\citet{2007astro.ph..2198D}).  Many of them have been applied in CMB
observations such as WMAP and Planck, and enabled high precision CMB
measurements (e.g. \citet{2015arXiv150205956P,2018arXiv180706208P} 
for a summary). Nevertheless, given the stringent requirement of 
accurate CMB measurement, there are still ongoing efforts to
improve existing methods or developing new methods (e.g. 
\citet{2018arXiv180700830U}). 

Here we report the {\bf ABS} method, which stands for the {\bf A}nalytical method of {\bf B}lind
{\bf S}eparation of CMB
from foregrounds. It can be treated as
post-processing on the matrix of cross bandpower between frequency bands, which are
heavily compressed products of the original (noisy) maps.  It works on any
single multipole bin.  Due to the fact 
that  CMB B-mode has a known (blackbody) frequency dependence, a set
of specific linear algebra  operations on this 
measured matrix automatically returns the bandpower $\mathcal{D}_{\rm  
  B}(\ell)$, the most important B-mode statistics. The measurement
procedure is completely fixed by  the measured matrix and 
survey specifications, with no assumptions on foregrounds. 
Since it does not rely on fitting procedures, it is numerically stable
and fast. This method was originally designed to solve the cross band power
  matrix, which is essentially identical to the matrix that  SMICA (spectral matching ICA,
  \citet{2003MNRAS.346.1089D,SMICA}) solves.  However,  after submitting the
manuscript, we were drawn the attention by CMB experts to its
similarity with ILC, despite of two different starting
points (power spectrum level versus map level). We verify that the ABS solution is
identical to the ILC solution in the limit of vanishing instrument
noise \citep{2008A&A...487..775V,
  2008PhRvD..78b3003S}. However, significant differences exist when the instrument
noise exists. We presents more detailed comparison to SMICA and ILC
(including ILC in the harmonic space, and the generalized needlet ILC
GNILC) in \S \ref{sec:conclusion}. 

This paper is organized as follows. In \S \ref{sec:method} we describe
the ABS method.  In \S \ref{sec:simulation} we generate simulated data
with various foreground components, CMB B-mode,  survey frequency
configurations and  instrumental noise. In \S \ref{sec:test} we test
the ABS method against these simulated data. In \S \ref{sec:bias} we derive
the necessary and sufficient survey conditions for unbiased CMB
measurement. In \S \ref{sec:conclusion} we discuss and compare ABS
with the  ILC and SMICA method. The 
appendix contains proof of a few key  results. 

\section{The ABS Method}
\label{sec:method}
The ABS method is motivated by the analytical solution of
$\mathcal{D}_{\rm B}$ derived under the ideal case of no instrument
noise (\S \ref{subsec:nonoise}). It is then extended to the case with instrument
noise (\S \ref{subsec:noise}). 

\subsection{The analytical solution for the case of no instrument
  noise}
\label{subsec:nonoise}
Our method works on $\mathcal{D}_{ij}(\ell)$, the
$N_f\times N_f$ matrix of cross bandpower between the $i$-th and $j$-th frequency
band. Here $\ell$ denotes the multipole bin.  $i,j=1, 2\cdots N_f$ and
$N_f$ is the total number of frequency bands.  
\be
\label{eqn:Dij}
\mathcal{D}_{ij}(\ell)= f^{\rm B}_if^{\rm B}_j \mathcal{D}_{\rm
  B}(\ell)+\mathcal{D}^{\rm fore}_{ij}(\ell)\ . 
\ee
Throughout this paper  we use the thermodynamic units, therefore $f^{\rm
  B}=1$. $\mathcal{D}^{\rm fore}_{ij}$ is the cross bandpower matrix of
foreground.  It has order $N_f$, but its rank $M$ depends on the number of independent
foreground components.  $\mathcal{D}_{\rm B}(\ell)$ is the band power
of CMB B-mode power spectrum, centered at multipole $\ell$. Our task is to solve Eq. \ref{eqn:Dij} for
$\mathcal{D}_{\rm B}(\ell)$, without assumptions of 
$\mathcal{D}^{\rm fore}_{ij}$. This may
appear as a mission impossible. However, due to the fact that CMB has
a blackbody spectrum, and the fact that there may be limited foreground
components in frequency space, Eq. \ref{eqn:Dij} may be indeed
solvable. We  are able to prove the following two key results.  
\bi
\item 
{\bf The solution to $\mathcal{D}_{\rm B}$
is unique, as long as $M<N_f$.}  The proof is given in the
appendix. A heuristic explanation is as follows. The matrix
$\mathcal{D}_{ij}$ has rank $M+1$. Subtracting 
$\mathcal{D}_{\rm B}f^{\rm B}_if_j^{\rm B}$, the new  matrix
$\mathcal{D}_{ij}-\mathcal{D}_{\rm B}f^{\rm B}_if_j^{\rm B}$ will have
rank $M$. Such reduction of $1$ in rank happens and only happens when
the trial value of $\mathcal{D}_{\rm B}$ exactly equals to its true
value.  This explains the existence and uniqueness of the
solution $\mathcal{D}_{\rm B}$. The above argument has assumed
  that the subspace extended by foreground eigenvectors contains no CMB
  direction. If this condition is violated, CMB can not be separated
  from foregrounds with spectral information alone.  Hereafter we will
work under this condition, unless otherwise specified. 
\item  {\bf The analytical solution exists, given by
\be 
\label{eqn:analyticalsolution}
\mathcal{D}_{\rm B}=\left(\sum_{\mu=1}^{M+1} G_\mu^2\lambda_\mu^{-1}\right)^{-1}\ . 
\ee }
\ei
Here the $\mu$-th eigenmode has eigenvector ${\bf
  E}^{(\mu)}$ and eigenvalue $\lambda_\mu$. We adopt the normalization ${\bf
  E}^{(\mu)}\cdot{\bf E}^{(\mu)}=1$. $G_\mu\equiv
{\bf f}^{\rm B}\cdot{\bf E}^{(\mu)}$.  We rank the eigenmodes with decreasing order in
$\lambda_\mu$. Since $\mathcal{D}_{ij}$ is positive definite,
$\lambda_\mu>0$.  The derivation of Eq. \ref{eqn:analyticalsolution},
based on the Sylvester's determinant theorem,  is given in the appendix. 

Eq. \ref{eqn:analyticalsolution} is not straightforward to
understand. However,  for the limiting case of $M\leq 2$, one can
solve for all eigenmodes analytically and verify
Eq. \ref{eqn:analyticalsolution} by brute-force. Another check,
although rather trivial, is that foreground components orthogonal to
the CMB signal in the frequency space indeed do not interfere the CMB
reconstruction. We emphasize that this ``orthogonality'' is a sufficient condition,
but not a necessary condition, for unbiased CMB reconstruction. The
sufficient and necessary condition is given before
Eq. \ref{eqn:analyticalsolution}. 

Eq. \ref{eqn:analyticalsolution} is not the only analytical expression
for $\mathcal{D}_{\rm B}$. A set of expression is as follows, 
\be 
\label{eqn:shift}
\mathcal{D}_{\rm B}=\left(\sum_{\mu=1}^{M+1}
  G_\mu^2\lambda_\mu^{-1}\right)^{-1}_{\mathcal{D}_{ij}+\mathcal{S} f^{\rm
    B}_if^{\rm B}_j}-\mathcal{S}\ . 
\ee
The shift parameter $\mathcal{S}$ is a free parameter. It shifts the input value
of CMB signal from $\mathcal{D}_{\rm B}$ to $\mathcal{D}_{\rm
  B}+\mathcal{S}$.  Namely, for the actual $\mathcal{D}_{ij}$ and an
arbitrary $\mathcal{S}$, we
  generate a new matrix $\mathcal{D}^{\mathcal{S}}_{ij}\equiv \mathcal{D}_{ij}+\mathcal{S} f^{\rm
    B}_if^{\rm B}_j$. We then obtain $\lambda_\mu$, ${\bf E}_\mu$ and
  $G_\mu$ with respect to this new matrix
  $\mathcal{D}^{\mathcal{S}}_{ij}$.   The first expression on the r.h.s. of
  Eq. \ref{eqn:shift} then returns $\mathcal{D}_{\rm
    B}+\mathcal{S}$. That is why we need to subtract $\mathcal{S}$ in
  Eq. \ref{eqn:shift} to
  obtain the correct $\mathcal{D}_{\rm B}$.  Eq. \ref{eqn:analyticalsolution} is a special case
of Eq. \ref{eqn:shift} with $\mathcal{S}=0$. If there are no instrument noises nor numerical
errors, Eq. \ref{eqn:analyticalsolution} \& \ref{eqn:shift} are
equivalent. However,  in reality Eq. \ref{eqn:shift}  with positive
$\mathcal{S}$ is more stable,
more accurate and therefore more  useful for the B-mode determination. 

\subsection{Extension to the case with instrument noise}
\label{subsec:noise}
We work under the condition that the ensemble average of the
instrument noise covariance matrix ($\langle \mathcal{D}_{ij}^{\rm inst}\rangle$) is
known. Notice that $\langle \mathcal{D}_{ij}^{\rm inst}\rangle$  may
have nonzero off-diagonal elements, due to correlated detector noises (e.g.\citet{2018arXiv180706207P,2018arXiv180706206P}
or atmosphere (e.g. \citet{2008ApJ...681..708P,2015ApJ...809...63E}). First we subtract this ensemble average. What remains in the
matrix is the residual instrument noise $\delta \mathcal{D}^{\rm
  inst}_{ij}\equiv \mathcal{D}_{ij}^{\rm inst}-\langle
\mathcal{D}_{ij}^{\rm inst}\rangle$, with zero
mean  ($\langle \delta \mathcal{D}^{\rm
 inst}_{ij}\rangle=0$).  The matrix that we deal with is then
\be
\label{eqn:Dobs}
\mathcal{D}^{\rm obs}_{ij}\equiv\mathcal{D}_{ij}+\delta
\mathcal{D}_{ij}^{\rm inst}\ .
\ee
Eq. \ref{eqn:analyticalsolution} \&  \ref{eqn:shift} can still be implemented in the data analysis,
with some modifications to account for instrument noise. 

{\bf Step 1}. First we need to deal with the varying residual noise across frequency bands.\footnote{Real surveys have other complexities.  The appendix \S
\ref{sec:appendixB} will show that the ABS method is still applicable with
the presence of masks and frequency dependent beams. }  In this case,
we should not treat each $\mathcal{D}_{ij}^{\rm obs}$ with equal
weight. The associated dispersion in each residual noise matrix element is  $\sigma^{\rm
  inst,2}_{\mathcal{D},ij}\equiv \langle (\delta \mathcal{D}^{\rm
 inst}_{ij})^2\rangle$.  Therefore we weigh $\mathcal{D}_{ij}^{\rm obs}$ by $\sqrt{\sigma^{\rm
    inst}_{\mathcal{D},ii}\sigma^{\rm inst}_{\mathcal{D},jj}}$,
\ba
\mathcal{D}_{ij}^{\rm obs} & \rightarrow &
\tilde{\mathcal{D}}_{ij}^{\rm obs}\equiv \frac{\mathcal{D}_{ij}^{\rm obs}}{\sqrt{\sigma^{\rm
    inst}_{\mathcal{D},ii}\sigma^{\rm inst}_{\mathcal{D},jj}}}\ .
\ea
By such normalization, the residual noise matrix in
$\tilde{\mathcal{D}}_{ij}^{\rm obs}$ has  
dispersion of $1$ in the diagonal elements. The dispersions of off-diagonal elements
depend on the residual noise property. For example, when the residual noise
is Gaussian,  the off-diagonal elements all have dispersion
$1/\sqrt{2}$. By doing so, we have downweighted the  band power
  measurements with large instrument noises. The ABS method also works
  without such weighting. But 
when the noise levels of different frequency bands differ
significantly, it will be far from optimal. 

 The ABS method applies to $\tilde{\mathcal{D}}_{ij}^{\rm obs}$, with the following operations,
\ba
f^{\rm B}_i &\rightarrow & \tilde{f}^{\rm B}_i\equiv \frac{f^{\rm B}_i }{\sqrt{\sigma^{\rm
    inst}_{\mathcal{D},ii}}}\ ,\no\\
G_{\mu} &\rightarrow & \tilde{G}_{\mu}\equiv \tilde{{\bf f}}^{\rm
  B}\cdot \tilde{{\bf E}}^{\mu}\ ,\ \lambda_\mu\rightarrow \tilde{\lambda}_\mu\ .
\ea
Here $\tilde{{\bf E}}^{\mu}$ is the $\mu$-th eigenvector of
$\tilde{\mathcal{D}}_{ij}^{\rm obs}$ and $\tilde{\lambda}_\mu$ is  the
eigenvalue.  The $i$-th diagonal element $\tilde{\mathcal{D}}_{ii}^{\rm
  obs}$ is the S/N of bandpower measurement of the $i$-th
frequency band.  Notice that the off-diagonal elements are not the S/N
of cross bandpower measurements, since we do not normalize by
$\sigma_{\mathcal{D},ij}^{\rm inst}$. The reason that we choose the
normalization $\sqrt{\sigma^{\rm
    inst}_{\mathcal{D},ii}\sigma^{\rm inst}_{\mathcal{D},jj}}$ is to
ensure the CMB contribution of the form   $f^{\rm B}_if^{\rm B}_j\mathcal{D}_{\rm B}$ or
$\tilde{f}^{\rm B}_i\tilde{f}^{\rm B}_j\mathcal{D}_{\rm B}$, which
have separable dependences on the $i$-th and $j$-th frequencies.

{\bf Step 2}. We also need  to deal with unphysical eigenmodes induced by
instrument noise.  With the presence of instrument noise, the rank of
$\tilde{\mathcal{D}}_{ij}^{\rm obs}$ will be $N_f$. The eigenmodes of
instrument noises have typical amplitude
$\sim 1$ and their distribution is symmetric. Therefore we must
exclude eigenmodes with negative eigenvalues. We should also exclude
eigenmodes with small eigenvalues. We choose the threshold
$\lambda_{\rm cut}\sim 1$.  We compute all
$N_f$ eigenmodes of $\tilde{\mathcal{D}}^{\rm obs}_{ij}$, and then measure $\mathcal{D}_{\rm B}$ from 
  Eq. \ref{eqn:shift}, but  only using eigenmodes with
$\tilde{\lambda}_\mu>\lambda_{\rm cut}$.  Namely,  the estimator of 
$\mathcal{D}_{\rm B}$ {\it with the presence of instrument noise} is 
\ba
\label{eqn:DB1}
\hat{\mathcal{D}}_{\rm B}=\left(\sum^{\tilde{\lambda}_\mu \geq \lambda_{\rm cut}} \tilde{G}_\mu^2\tilde{\lambda}_\mu^{-1}\right)_{\tilde{\mathcal{D}}_{ij}^{\rm obs}+\mathcal{S} \times \tilde{f}^{\rm
  B}_i\tilde{f}^{\rm B}_j}^{-1}-\mathcal{S}\ . 
\ea

{\bf Step 3}. In this step we carry out a convergence
test/self-calibration procedure to
determine a suitable choice of $\mathcal{S}$ and then use it to obtain
$\mathcal{D}_{\rm B}$. 
$\mathcal{S}$ changes the distribution of physical eigenmodes.  Larger positive $\mathcal{S}$ makes the matrix operations more
stable and the impact of instrument noise weaker. By increasing $\mathcal{S}$ and
finding the converged value of  $\hat{\mathcal{D}}_{\rm
  B}$, we obtain a more reliable measure of $\mathcal{D}_{\rm
  B}$. We emphasize two points. First, both $\mathcal{S}$ and
$\mathcal{D}_{\rm B}$ are self-determined from the data and no extra
uncertainties are introduced in this step. Second, this step is also
necessary to pass the null test detailed later.

\section{Simulated observations for tests}
\label{sec:simulation}
Next we test the ABS method on simulated $\mathcal{D}^{\rm obs}_{ij}$
with a variety of foregrounds, instrument noise and survey frequency
configurations.  

\bfi{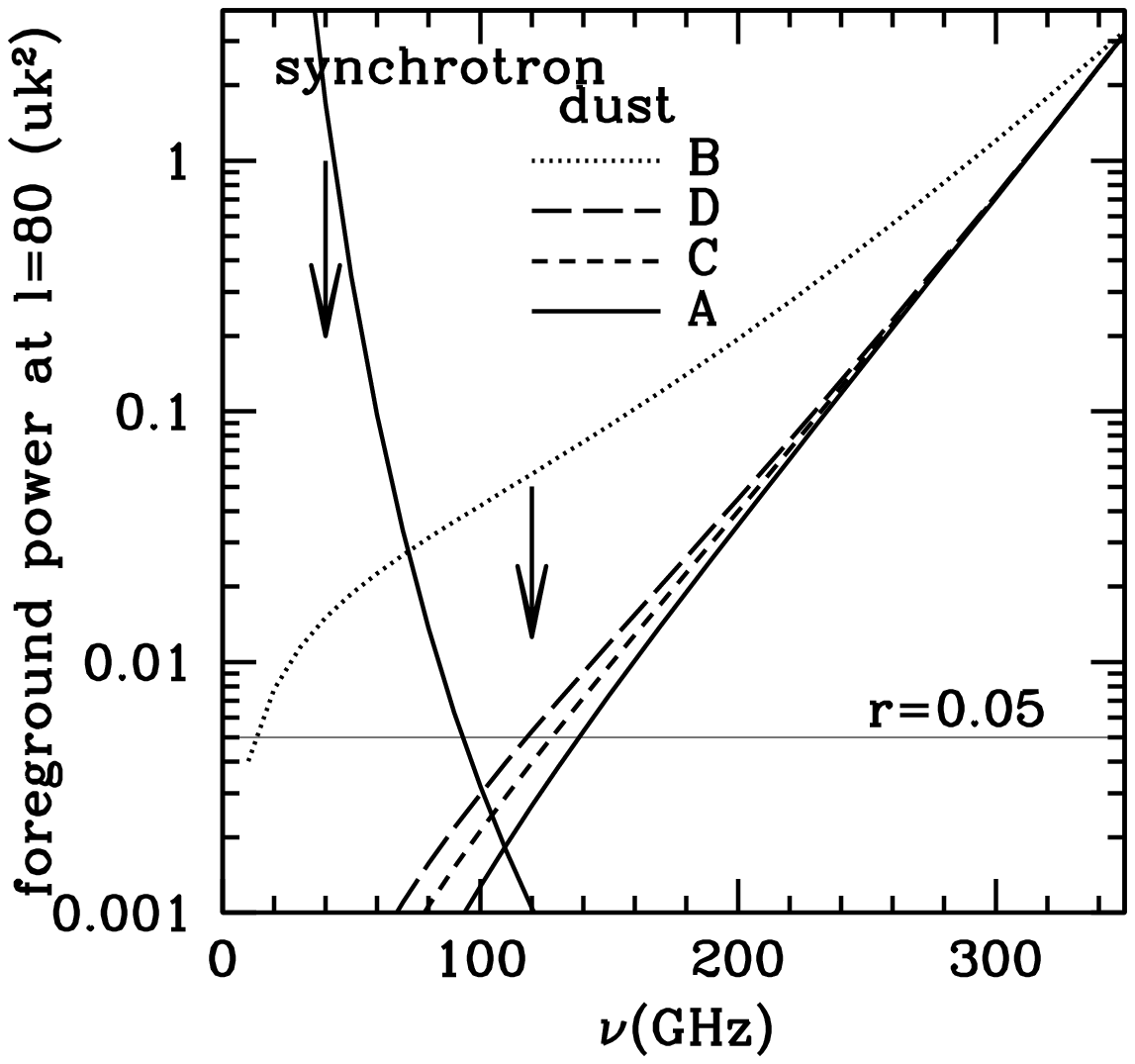}
\caption{The foreground models for simulated observations to test the
  ABS method. The $4$ models share identical synchrotron foreground, but different dust model
  parameters. Case C and D differ from case A and B by  significant decorrelation between dust
  foregrounds at different frequencies. Case B and D have exaggerated dust
  contamination and are served to test the generality of our method. \label{fig:f}}
\efi

\subsection{Foreground specifications}
 For foregrounds, we specify 
\ba
\label{eqn:foreground}
\mathcal{D}^{\rm fore}_{ij}=\sum_{\alpha=1}^{M} f^{(\alpha)}_if^{\alpha)}_j
\mathcal{D}_\alpha\ .
\ea
 $f^{(\alpha)}_i\equiv f^{(\alpha)}(\nu_i)$ is the frequency dependence of the $\alpha$-th
foreground component and $\mathcal{D}_\alpha$ is the bandpower
amplitude. 
Throughout the paper, we include two polarized foregrounds (galactic dust and galactic
synchrotron).  When we consider decorrelation between galactic dust at different
frequency, we need at least two  $f^{\alpha}(\nu)$ to describe
dust alone.  Therefore $M=2$ if no decorrelation and 
$M\geq 3$ when decorrelation exists.  

We  consider   four foreground models (case A, B, C, D, Fig. \ref{fig:f}). They all 
share the same synchrotron foreground, {\it but different dust
  foregrounds}. For synchrotron, 
\ba
f^{\rm syn}(\nu)\propto \nu^{-\beta_{\rm syn}/2}\frac{(e^x-1)^2}{e^xx^4}\ ,\ \mathcal{D}_{\rm
  syn}\propto \ell^{-0.6}\ .
\ea
Here $\beta_{\rm syn}=3.3$ is the frequency index and $x\equiv
h\nu/(k_BT_{\rm CMB})$. The bandpower is normalized as $3\times
10^{-4} \mu {\rm K}^2$, at $\nu=150$ GHz and $\ell=80$. This is the observationally allowed
upper limit in the BICEP2 sky \citep{2015PhRvL.114j1301B}.

For the galactic dust foreground, we adopt  \citep{2016A&A...586A.133P}
\ba
f^{\rm dust}(\nu)\propto 
\frac{x^{\beta_{\rm d}} (e^x-1)^2}{xe^x(e^{xT_{\rm CMB}/T_{\rm d}}-1)}\ ,\ \mathcal{D}_{\rm dust}\propto \ell^{-0.42}\ .
\ea
To account for the recently detected decorrelation between different Planck
frequency bands \citep{2016arXiv160607335P}, we adopt a simple model of spatially stochastic variation in the dust index $\beta_{\rm
  d}$\citep{2016arXiv160607335P}. It induces a new component in
$\mathcal{D}_{ij}(\ell)$,
\be
f^{\rm S}(\nu)= f^{\rm dust}(\nu)\times \ln (\nu/\nu_0)\ ,\ 
\mathcal{D}_{\rm S}=\mathcal{D}_{\rm dust}A_{\rm S}\ .
\ee
Here we adopt $\nu_0=353$ GHz. $A_{\rm S}\propto \langle \delta \beta_{\rm
  d}^2\rangle$ is a free parameter to control the level of
decorrelation. When this stochastic component is subdominant, the cross correlation
coefficient between dust in $i$-th and $j$-th bands is
$\mathcal{R}^{\rm BB}_{\nu_i\nu_j}\simeq
1-\frac{1}{2}A_{\rm S}(\ln(\nu_i/\nu_j))^2$. The overall bandpower is
normalized as $3.5\mu {\rm K}^2$ at $\ell=80$ and $353$
GHz \citep{2015PhRvL.114j1301B}. 

We adopt 4 cases of dust parameters,  $(\beta_{\rm d}, T_{\rm d},A_{\rm
  S})$ = $(1.59, 19.6, 0.0)$, $(0.5, 10, 0.0)$,
$(1.59, 19.6, 0.42)$, $(1.59, 19.6, 0.84)$. Case A is the best fit
of Planck \citep{2016A&A...586A.133P}.  Case B  has
  a factor of $10$ more dust contamination at $100$-$150$ GHz than
  case A, and also a much flatter spectrum. Case C has dust decorrelation 
  between frequency bands, reproducing the Planck finding
  of  $\mathcal{R}^{BB}_{353,217}=0.95$
  \citep{2016arXiv160607335P}. Case D has  unrealistically large
  decorrelation (e.g. $\mathcal{R}^{BB}_{353,150}=0.7$).

\begin{table}
\caption{We test our ABS method against various CMB frequency
  configurations and instrumental noise.  $\sigma_{\mathcal{D}}^{\rm inst}$ is the r.m.s. error
  in the bandpower measurement caused by instrumental noise. }
\begin{tabular}{c|c|c} \\
Labels&frequency/GHz & $\sigma_{\mathcal{D}}^{\rm inst}$/$\mu{\rm K}^2$ \\ \hline\hline
{\bf F}0 & 30, 70, 100, 150, 217 \& 353 &  \\
{\bf F}1 & 95, 150, 220 \& 270 & \\ 
{\bf F}2 & 35, 95, 150, 220 \& 270 & \\ 
{\bf F}3 & 35, 95, 150, 220, 270 \& 353 &$( 10^{-5}, 10^{-2})$\\
{\bf F}4 & 30, 36, 43, 51, 62, 75, 90,105, 135 & \\
&160, 185,200, 220, 265, 300 \& 320 & \\ \hline\hline
\end{tabular}
\label{table:S}
\end{table}

\subsection{Frequency configurations}
Frequency configuration is crucial for foreground removal. We
consider five configurations ({\bf F}0-{\bf F}4), shown in Table
\ref{table:S}. 
\bi
\item {\bf F}0 is the
fiducial one, with $6$ bands centered at 30, 70, 100, 150, 217 \& 353
GHz. This configuration is similar to Planck. It has a wide frequency
coverage, good for both synchrotron and dust foreground removal. 
\item  {\bf F}1 has  4 bands at 95, 150, 220 \& 270 GHz (Keck
array-like, \citet{BICEP3}). A major difference of {\bf F}1 to {\bf
  F}0 is the lack of low frequency bands and hence limited capability
of synchrotron foreground identification and removal.  
\item  {\bf F}2 adds a $35$ GHz bands to {\bf F}1 (BICEP
array-like, \citet{BICEP3}).  This is to test the gain adding a low
frequency band. 
\item {\bf F}3 further adds a 353 GHz band to {\bf F}2. This turns out
  to be important for dust foreground removal when decorrelation in
  dust foreground exists. 
\item  {\bf F}4 has 16
bands between $30$ GHz and $320$ GHz. This is basically the frequency
configuration of PRISM \citep{PRISM}, expect that PRISM also has 
higher frequency bands. Other proposed space missions such as CORE,
PIXIE and LiteBIRD have similar configurations. 
\ei

\subsection{B-mode signal and physical eigenmodes}
For the CMB signal,  we focus on $\ell=80$ around the recombination bump. The
fiducial $\mathcal{D}_{\rm B}=5\times 10^{-3}\mu {\rm K}^2$,
corresponding to the sum of $r=0.05$ and  the lensing B-mode. We also  
consider $\mathcal{D}_{\rm B}=2\times 10^{-3}\mu {\rm K}^2$ in which
the lensing B-mode dominates.  We further test around $\ell=5$ of
the reionization bump, with the choices of $\mathcal{D}_{\rm B}=1$,
$2\times 10^{-3}\mu {\rm K}^2$. 

\bfi{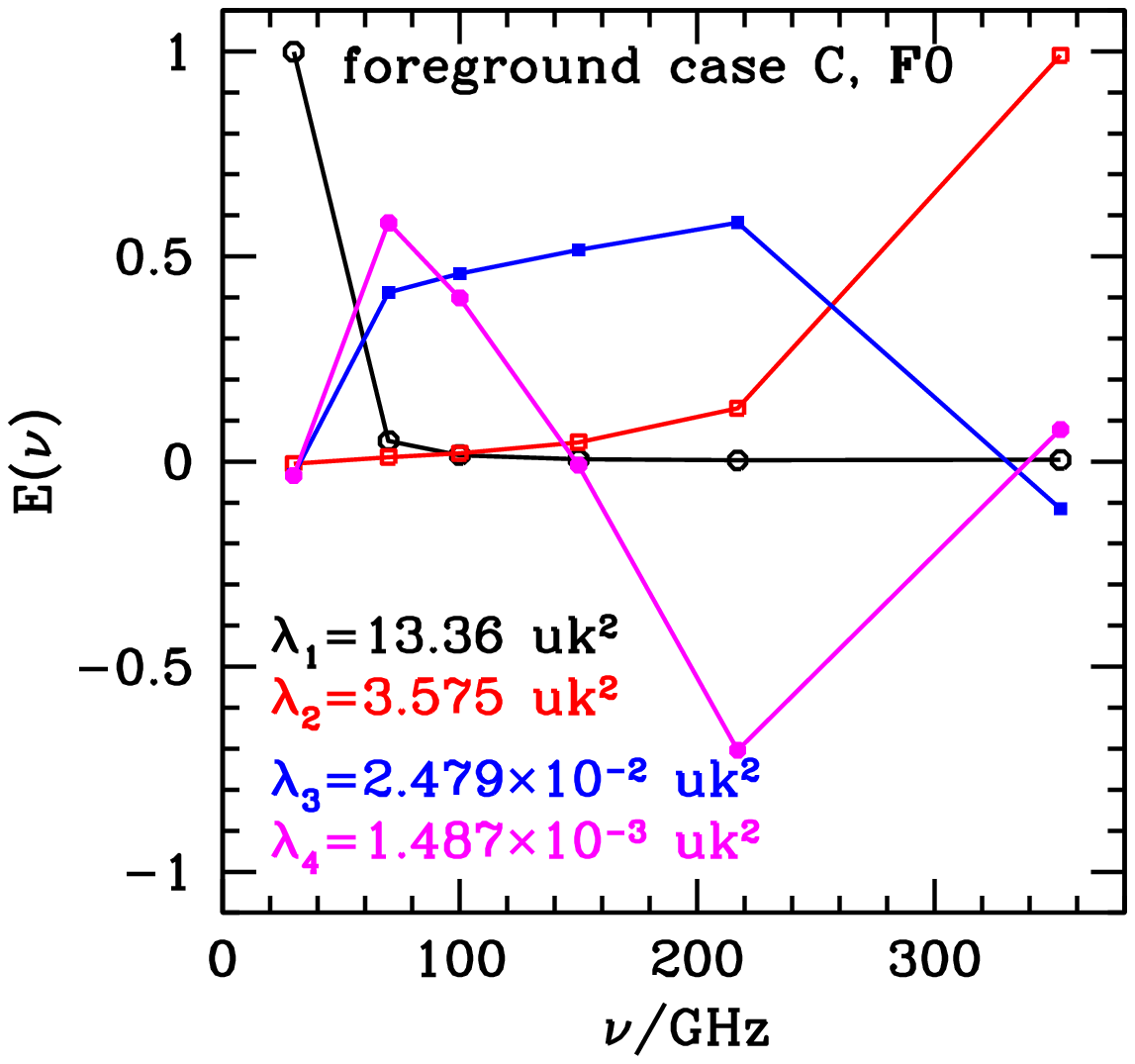}
\caption{The eigenvectors and eigenvalues for foreground case C and
  frequency configuration {\bf F}0, with
  $\mathcal{D}_{\rm B}=5\times 10^{-3} \mu{\rm K}^2$ and centered at
  $\ell=80$. Due to significant decorrelation of thermal dust
  foreground, case C has $4$ eigenmodes.  (1) The first eigenmode (open circle) is dominated by
  synchrotron foreground, with eigenvalue $\lambda_1$ essentially its band
  power at $30$ GHz.  Since the synchrotron model is actually the
  observational upper limit, this eigenmode may be less
  significant in reality. (2) The second eigenmode (open square) is
  dominated by dust emission, with $\lambda_2$ essentially the dust
  emission band power at $353$ GHz. (3) The third eigenmode (filled
  square) is dominated by CMB B-mode. (4) The fourth
  eigenmode (filled circle) is a mixture of all foreground and CMB
  components. Although it is subdominant, it is important for unbiased
  CMB measurement. \label{fig:e}}
\efi

\bfi{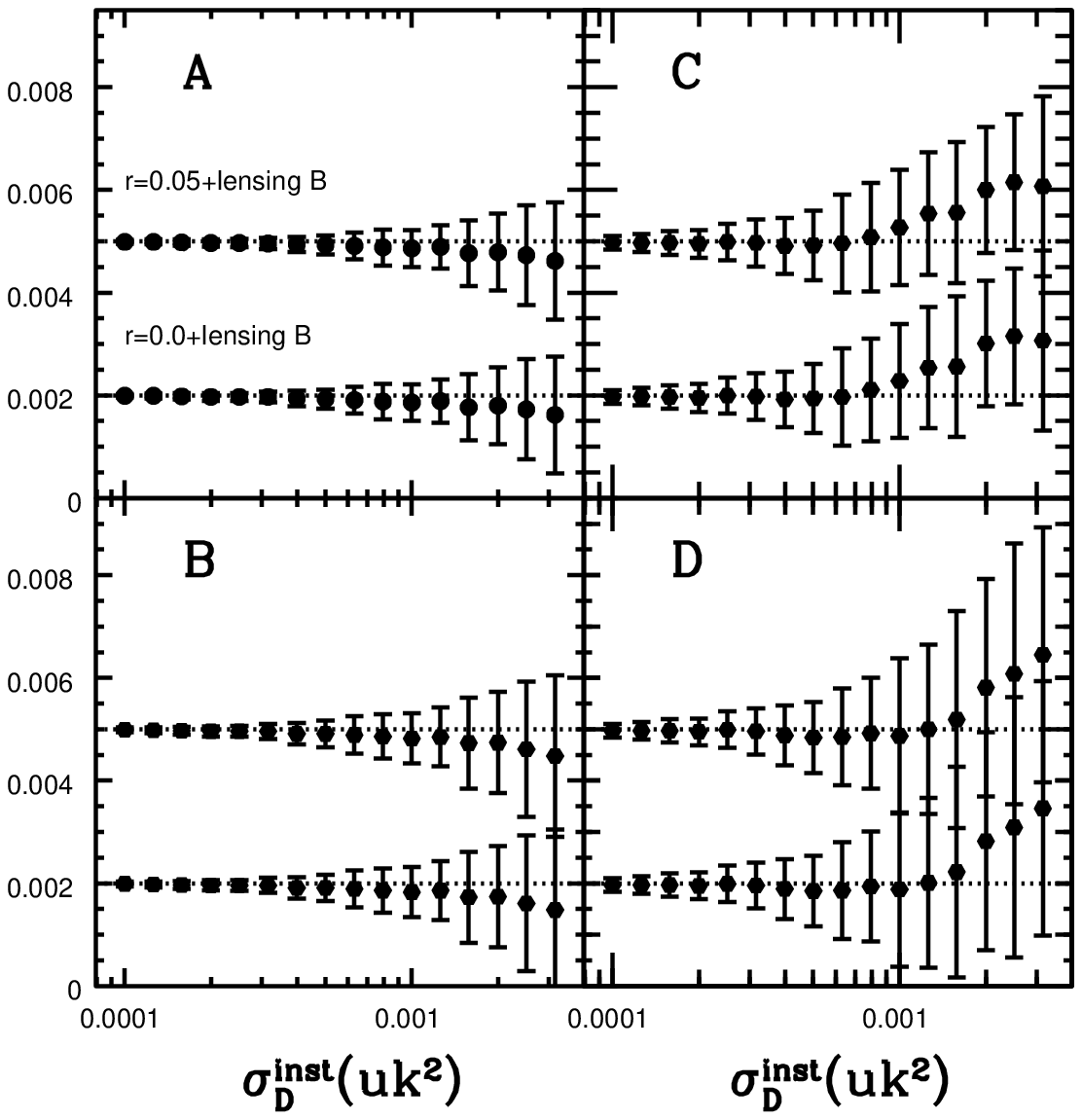}
\caption{Tests for the {\bf F}0 survey configuration and $4$ cases of
  foregrounds. The $y$-axis is the bandpower in unit of $\muk^2$. Dot lines are the  input 
  B-mode. Points are the ABS output, as a function of bandpower
  measurement error $\sigma_{\mathcal{D}}^{\rm inst}$.  The error bars are estimated using $200$
  realizations of instrument noise.  The input
  $\mathcal{D}_{\rm B}$ is recovered unbiasedly. We emphasize that the adopted foreground cases
  are to generate simulated observational data. Our ABS method
  assumes nothing about these foregrounds.  \label{fig:ABCD80}}
\efi

The eigenmodes of $\mathcal{D}_{ij}$ depend on foregrounds, CMB signal
and observational frequency configuration.  Two useful relations to
understand these eigenmodes are
\ba
\sum_{\alpha=1}^{M+1} \lambda_\alpha&=&{\rm Tr}\mathcal{D}_{ij}=\sum_{i=1}^{N_f}
\mathcal{D}_{ii}\ ,\no\\
\sum_{\alpha=1}^{M+1} \lambda^2_\alpha&=&\sum_{ij}\mathcal{D}^2_{ij}\ .
\ea
Fig. \ref{fig:e} shows the eigenmodes for foreground model C, $\mathcal{D}_{\rm B}=5\times
10^{-3}\mu {\rm K}^2$, and frequency configuration {\bf F}0. It has
$4$ eigenmodes. The first two are essentially synchron and dust
foreground, respectively. These can be seen from their frequency
dependences (the shapes of eigenvectors). Furthermore, $\lambda_1\simeq \sum_i
\mathcal{D}^{\rm syn}_{ii}\sim \mathcal{D}^{\rm syn}_{11}$, $\lambda_2\simeq \sum_i
\mathcal{D}^{\rm dust}_{ii}\sim \mathcal{D}^{\rm dust}_{66}$.  The
third one is dominated by CMB, $\lambda_3\simeq 5 \mathcal{D}_{\rm
  B}$. It is close to $N_f\mathcal{D}_{\rm B}=6\mathcal{D}_{\rm B}$, the limit of
pure CMB B-mode. For the same reason,
it contains non-negligible contamination from foregrounds. The fourth
eigenmode is a mixture of CMB and foregrounds, with a frequency
dependence resembling none of CMB and foregrounds. This eigenmode is
also important for CMB measurement, as will be shown later. 

\subsection{Instrument noise specifications}
To generate simulated $\mathcal{D}^{\rm obs}_{ij}$, we approximate $\delta 
\mathcal{D}^{\rm inst}_{ij}$ as Gaussian random fields with dispersion
$\sigma^{\rm inst}_{\mathcal{D},ij}$ and  $\sigma^{\rm
  inst,2}_{\mathcal{D},ij}=\sigma^{\rm
    inst}_{\mathcal{D},i}\sigma^{\rm
    inst}_{\mathcal{D},j}(1+\delta_{ij})/2$.  For brevity, we
assume $\sigma^{\rm inst}_{\mathcal{D},11}=\sigma^{\rm inst}_{\mathcal{D},22}=\cdots=\sigma^{\rm
  inst}_{\mathcal{D}}$ ($i=1,\cdots,N_f$). Therefore $\mathcal{D}^{\rm
  obs}_{ij}$ and $\tilde{\mathcal{D}}^{\rm obs}_{ij}$ only differ by a
uniform normalization. This allows us to work directly on $\mathcal{D}^{\rm
  obs}_{ij}$, whose physical meaning is clearer than $\tilde{\mathcal{D}}^{\rm
  obs}_{ij}$.  Notice that the off-diagonal elements have  smaller
dispersion ($\sigma^{\rm  inst}_{\mathcal{D}}/\sqrt{2}$). 

Reducing instrument noise  is a key task  in CMB polarization experiments.  BICEP2/Keck has reached $\sigma^{\rm inst}_{\mathcal{D}}\sim 10^{-3}\mu
{\rm K}^2$ \citep{2015PhRvL.114j1301B}. Future experiments can go well below $10^{-4}\mu
{\rm K}^2$. For example, planned ground CMB-S4 projects \citep{CMB-S4} will have two
orders of magnitude more detectors than BICEP2 ($\sim 5\times 10^5$)
and therefore a factor of $10$ reduction in instrument noise.  PRISM \citep{PRISM} has typical noise $\sim
  70\mu{\rm K}$/detector/arcmin$^2$, $\sim 200$
  detectors per band, 
$\sigma^{\rm inst}_{\mathcal{D}}\simeq3.7\times10^{-5}\mu{\rm K}^2(\ell/\Delta
    \ell)^{1/2}(0.5/f_{\rm
      sky})^{1/2}(\ell/80)$. Here $\Delta \ell$ is the width of
    multipole bin and $f_{\rm sky}$ is the fractional sky
    coverage. Other experiments such as CORE, EPIC and LiteBIRD have
    similar sensitivity.  We consider a wide range of $\sigma^{\rm inst}_{\mathcal{D}}\in( 10^{-5},
10^{-2})$$\mu{\rm K}^2$ to include all these possibilities.

\bfi{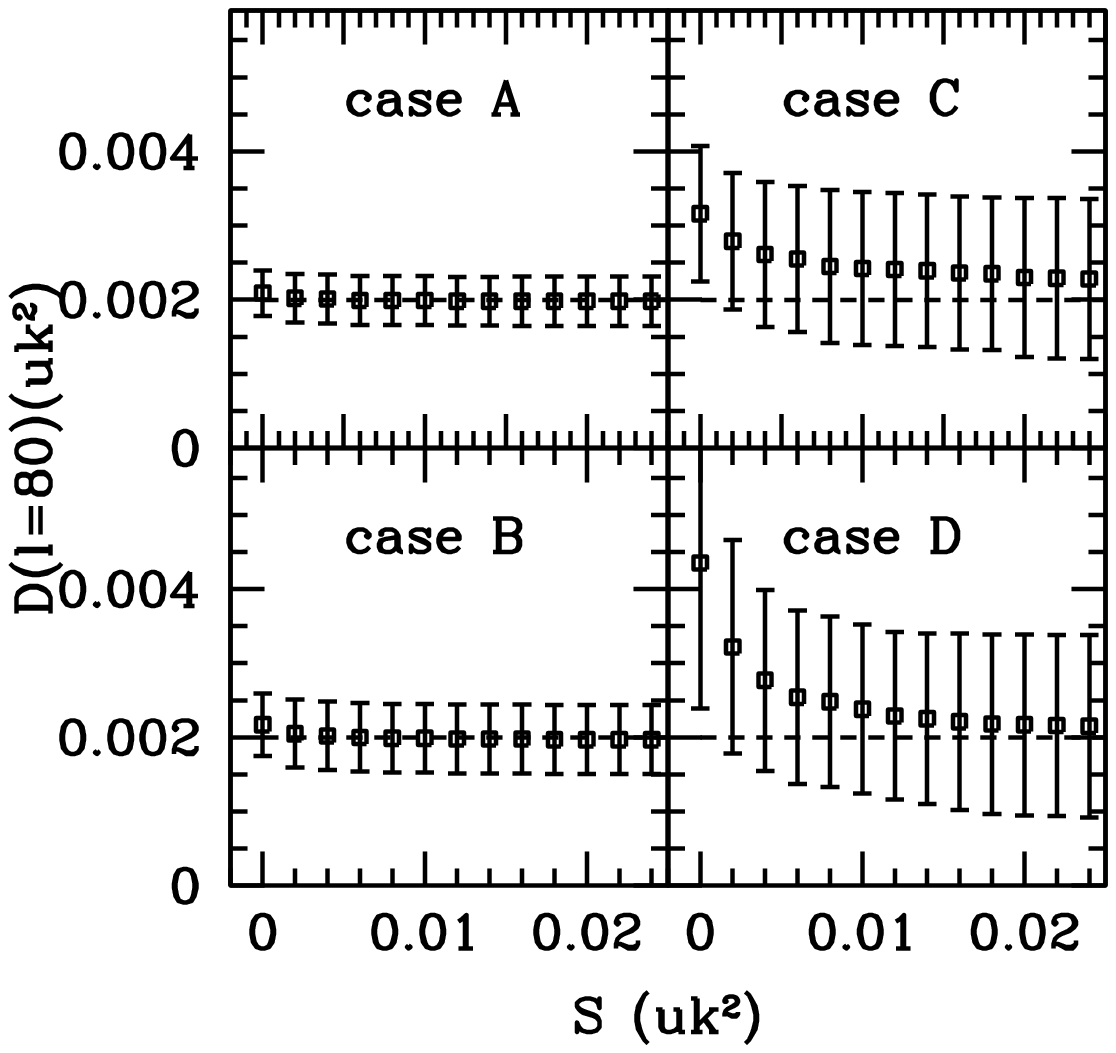}
\caption{The convergence test and self-calibration process. We test
  the output $\mathcal{D}_{\rm B}$ as the function of $\mathcal{S}$, for $\mathcal{D}_{\rm B}=2\times 10^{-3}\muk^2$
and $\sigma_{\mathcal{D}}^{\rm inst}=10^{-3}\muk^2$. By choosing
sufficiently large $\mathcal{S}$, systematic error in the reconstruction
can indeed be alleviated.  \label{fig:SA}}
\efi

\section{Testing the ABS method }
\label{sec:test}
Fig. \ref{fig:ABCD80} shows the test result for the {\bf F}0 frequency
configuration at $\ell\sim 80$. Throughout the paper, we fix the cut
$\lambda_{\rm cut}=1/2$. Whether this choice of $\lambda_{\rm cut}$ is
optimal is an open
question for future investigation.  The statistical error, for each
noise level, foreground and signal,  is estimated
using $200$ realizations of instrument noises (but identical CMB and foregrounds). For all investigated foregrounds,
signal and 
noise levels, our method faithfully extracts the input B-mode.    It is unbiased
even for high level of instrument noise $\sigma_{\mathcal{D}}^{\rm
  inst}\sim \mathcal{D}_{\rm B}$.

\subsection{Convergence test and self-calibration}
\label{subsec:convergence}
The results in Fig. \ref{fig:ABCD80} adopt $\mathcal{S}=20
\sigma_{\mathcal{D}}^{\rm  inst}$. The CMB reconstruction with such
choice of $\mathcal{S}$ is excellent. The choice of
$\mathcal{S}$ is not important when $\sigma_{\mathcal{D}}^{\rm
  inst}\ll \mathcal{D}_{\rm B}$. However, it will become important
when $\sigma_{\mathcal{D}}^{\rm inst}\sim 
\mathcal{D}_{\rm B}$. For $\mathcal{D}_{\rm B}=2\times 10^{-3}\muk^2$
and $\sigma_{\mathcal{D}}^{\rm inst}=10^{-3}\muk^2$, Fig. \ref{fig:SA}
shows the ABS output indeed varies with the choice of
$\mathcal{S}$. How shall we fix this degree of freedom?  We argue that
a nearly optimal choice of $\mathcal{S}$ can be  self-determined by the data, through
the step 3 of the ABS method (\S \ref{subsec:noise}).  By adding
$\mathcal{S}f_i^{\rm B}f_j^{\rm B}$ to the observed $\mathcal{D}^{\rm
  obs}_{ij}$ and running ABS with increasing $\mathcal{S}$, we find that the output converges
when $\mathcal{S}/\sigma_{\mathcal{D}}^{\rm  inst}\ga 10$
(Fig. \ref{fig:SA}). Furthermore,  when the ABS output converges, the bias in $\mathcal{D}_{\rm B}$ also
vanishes (Fig. \ref{fig:SA}).  For example, when $\mathcal{S}=0$,
the systematic bias  is  greater than $1\sigma$ for  case
C and D (Fig. \ref{fig:SA}). But when $\mathcal{S}\ga 10
\sigma_{\mathcal{D}}^{\rm  inst}$, the bias becomes statistically
insignificant.  For these reasons,  we will adopt $\mathcal{S}=20
\sigma_{\mathcal{D}}^{\rm  inst}$ throughout the paper, unless
otherwise specified.  Given the good performance of ABS on a variety
of foreground, CMB and survey specifications tested in the paper, we
expect that this choice of $\mathcal{S}$ is close to
optimal. Nevertheless, whether this is optimal and whether we can fix
the optimal $\mathcal{S}$ through the data alone are issues for further investigation. 

\bfi{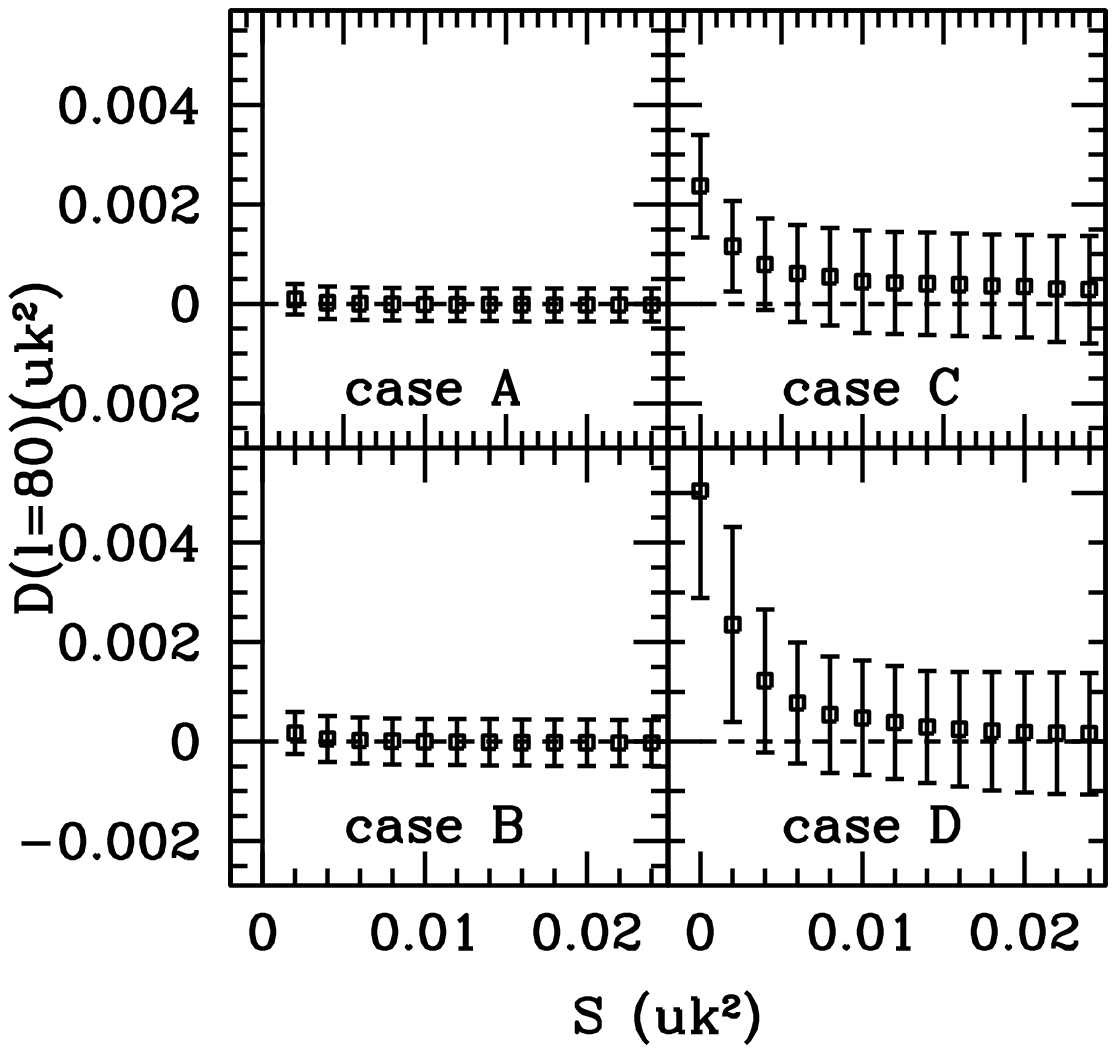}
\caption{The null test result for the noise level
  $\sigma_{\mathcal{D}}^{\rm inst}=10^{-3}\muk^2$. We set the signal as zero and check the ABS
  output. The $\mathcal{S}=0$ version of the ABS method fails the null
  test since by design it always returns positive value. But $\mathcal{S}$ which passes the convergence test (Fig. \ref{fig:SA}) automatically passes the null
  test.    \label{fig:null}}
\efi

\subsection{Null test}
We also carry out a null test of the ABS method by setting the
input signal zero. The $\mathcal{S}=0$ version of ABS
(Eq. \ref{eqn:DB1}) fails the null test since
it always returns positive value. Furthermore, the output result can be very
unstable (e.g. leftmost data points of
Fig. \ref{fig:null}). Fortunately with $\mathcal{S}\sim 10\sigma^{\rm inst}_{\mathcal{D}}$ that
can pass the convergence test,  the null test is also passed. This
again demonstrates that step 3 of the ABS method is necessary. We
address here that the choice of $\mathcal{S}$ does not induce extra
uncertainty in the CMB measurement, because it is completely fixed by
the data itself through the convergence test in \S
\ref{subsec:convergence}. 

\subsection{Statistical errors}
Fig. \ref{fig:error} plots the statistical error of the estimated
$\mathcal{D}_{\rm B}$ as a function of instrument noise
$\sigma_{\mathcal{D}}^{\rm inst}$. Roughly speaking,  the
statistical error $\sigma_{\rm B} \propto
\sigma_{\mathcal{D}}^{\rm inst}$. This is what we expect from our
analytical prediction.  Fig. \ref{fig:error} also compares $\sigma_{\rm B}$  with $\sigma_{\rm
  min}\equiv \sigma^{\rm inst}_{\mathcal{D}}/\sqrt{N_f(N_f+1)/2}$. The
later is  a lower bound of statistical error. It corresponds to the
limit of no foreground contaminations in which we can simply average
over all cross correlation measurements. The presence of foreground
enlarges the statistical error, by a factor of $\sim 2$ for case A/B
and a factor of $\sim 6$ for case C/D. 

\bfi{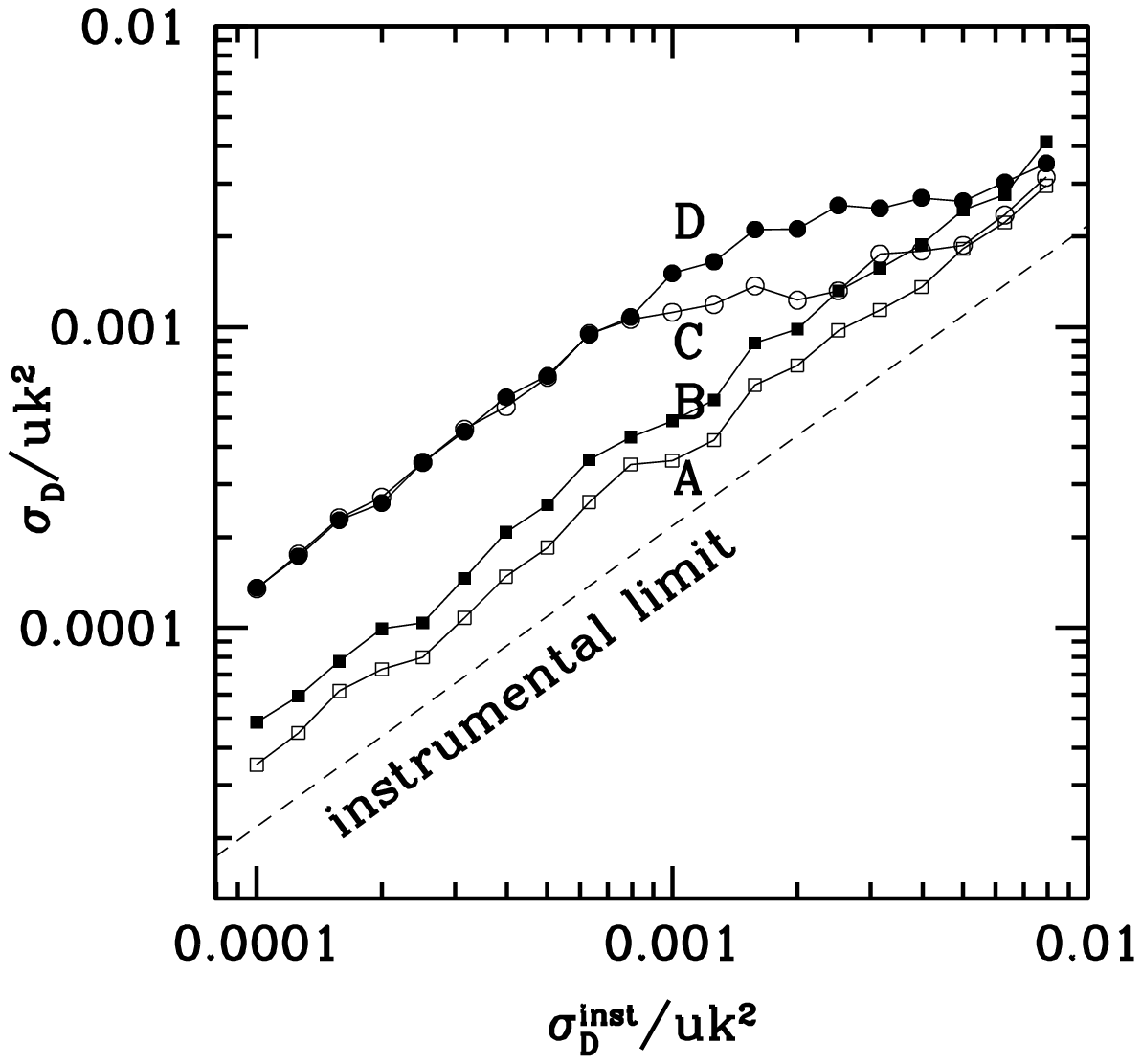}
\caption{The dependence of the statistical error
  $\sigma_{\mathcal{D}}$ on the instrument bandpower error  $\sigma_{\mathcal{D}}^{\rm inst}$ per band. $\sigma_{\mathcal{D}}$
  at $\ell\sim 80$ and survey configuration ``{\bf F}0'' is shown for the foreground cases of A, B, C and D,
  respectively.  The dash line is the instrumental limit
  $\sigma_{\mathcal{D}}^{\rm inst}/\sqrt{N_f(N_f+1)/2}$ which can only
  be achieved when foregrounds are negligible. \label{fig:error}}
\efi

\subsection{Insensitivity to foregrounds}
The above results also show that the recovery of  B-mode by ABS is insensitive to the  overall amplitude and spectral shape of galactic
foregrounds. For example, case B has a factor of $\sim 10$ larger dust contamination
at $\sim 150$ GHz band than case A. It also has a much flatter
spectrum. Both would severely degrade the CMB extraction. However,
the performance is almost as good as case A, without statistically
significant bias. The only major difference is that the  statistical
error is about  $40\%$ larger (Fig. \ref{fig:error}). 

Furthermore, our ABS method also works when decorrelation of
foregrounds at different frequencies exists (case C \& D,
Fig. \ref{fig:ABCD80}).  Case C \& D have one more dust component, so 
one can not simply scale from high frequency maps to low frequency maps to
remove dust foreground.  Our method nevertheless recovers the input
B-mode, robustly and blindly.  This demonstrates the advantage that the ABS method needs no
assumptions on the number of independent foreground components. 

Fig. \ref{fig:ABCD5} shows the  test results at $\ell\sim 5$ around the
reionization bump (Fig. \ref{fig:ABCD5}). Again the ABS successfully
recovers the input B-mode. The signal, foregrounds and instrument
noises at $\ell\sim 5$ are very different to that at $\ell\sim
80$. The synchrotron and dust foregrounds are a factor of $5$ and $3$
larger, respectively.  The B-mode signal is dominated by primordial
gravitational wave B-mode, with an amplitude $\la 2\times
10^{-3}\muk^2$. Therefore the overal B-mode signal to foreground ratio
is a factor of $\sim 10$ smaller than that at $\ell\sim 80$.  Success
of ABS for the $\ell\sim 5$ then further demonstrates  its
insensitivity to foreground properties.

\bfi{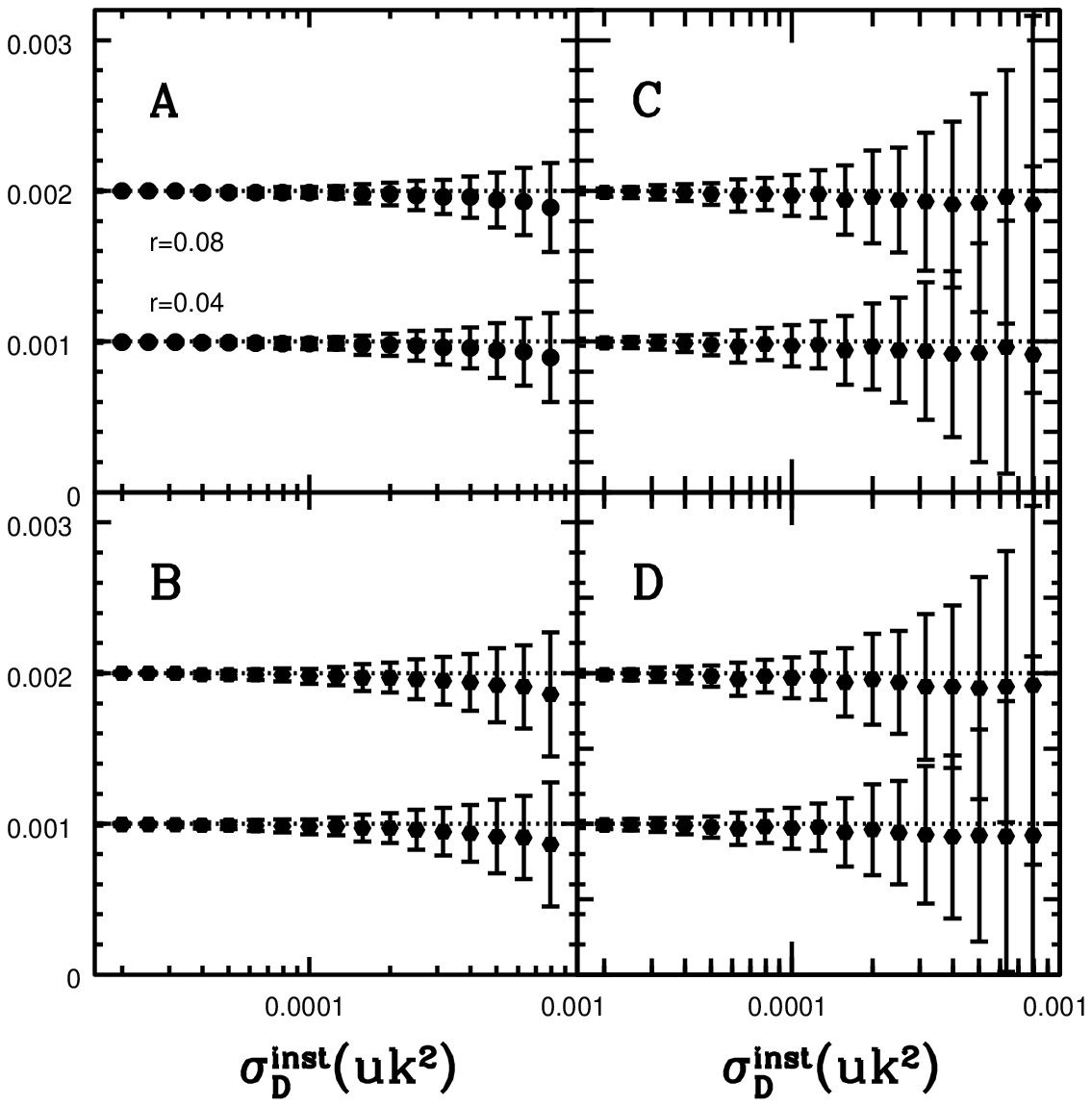}
\caption{Similar to Fig. \ref{fig:ABCD80}, but for $\ell=5$ around the
  reionization bump. The $y$-axis is the bandpower in unit of $\muk^2$ Notice that, due to smaller B-mode at this
  scale and smaller instrumental noise (scales as $\ell^{-1}$), we use different range of
  $\sigma_{\mathcal{D}}$ to that in Fig. \ref{fig:ABCD80}. \label{fig:ABCD5}}
\efi

\section{Survey requirements for unbiased measurement}
\label{sec:bias}
The success of ABS against the variety of  foreground models, CMB
signal and instrument noise levels investigated above is 
encouraging. Nonetheless, certain survey requirements have to be
satisfied to achieve unbiased CMB measurement. If a survey is lack of
necessary frequency coverage or is lack 
of necessary sensitivity, it may fail to correctly identify one or
more foreground components. If such foregrounds are not orthogonal to
CMB in frequency space, they will then  lead to biased CMB
estimation.  The ABS method
provides a specific diagnostic.

\bfi{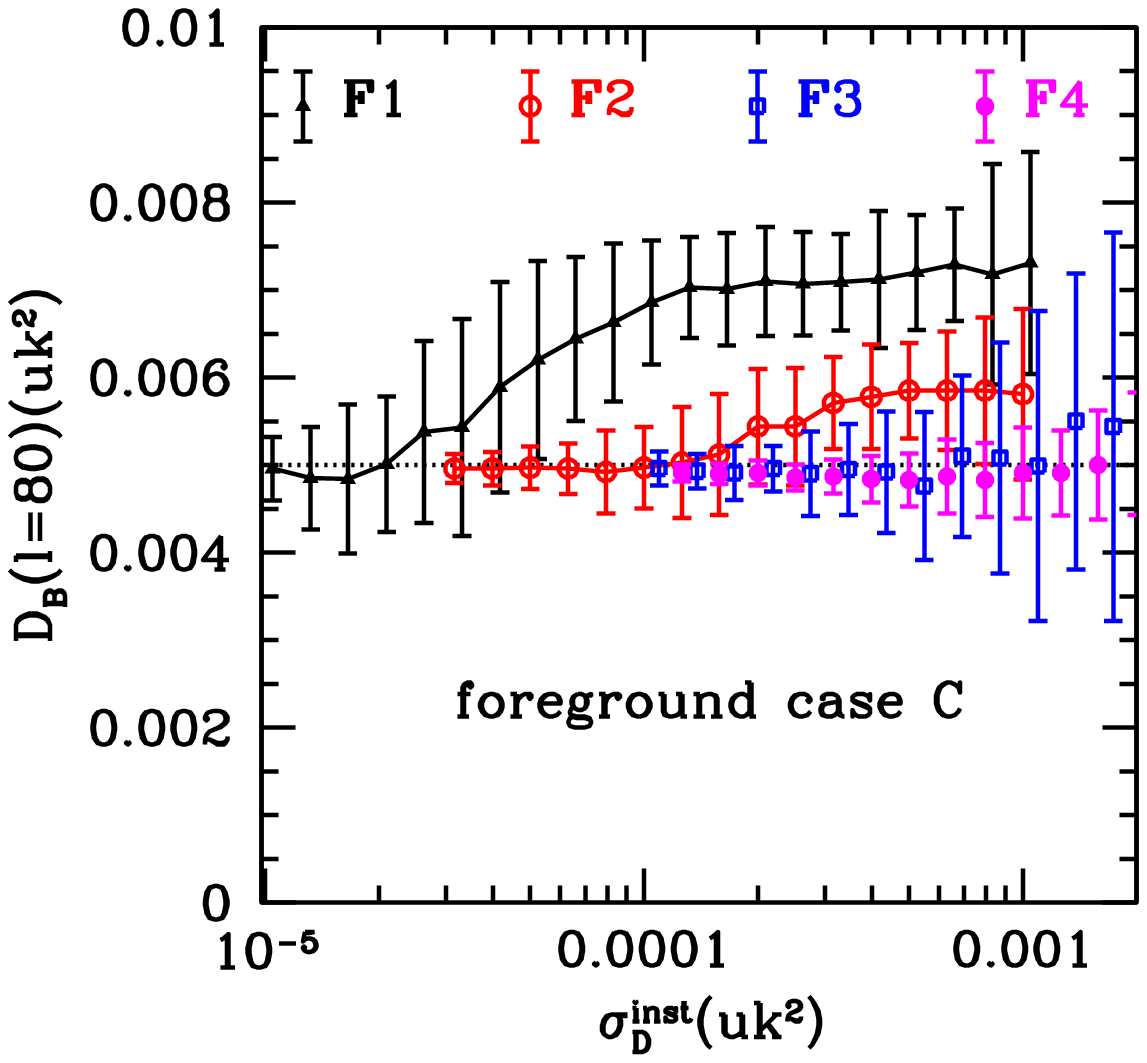}
\caption{The impact of frequency coverage on CMB signal
  extraction. Incomplete frequency coverage (e.g. {\bf F}1) and insufficient
  sensitivity cause failure in identifying certain
  eigenmodes significant for B-mode measurement
  (Fig. \ref{fig:ev}). This survey limitation causes bias in
  $\mathcal{D}_{\rm B}$. \label{fig:S1234}}
\efi
\subsection{Bias induced by survey limitations}
 We demonstrate this point with the {\bf F}1-4 frequency
configuration. It turns out that ABS still remains unbiased for model A and
B, for all relevant noise levels. Therefore for brevity we only show
the tests results for foreground case C (Fig. \ref{fig:S1234}). 

The foreground case C for testing has a large synchrotron
component, together with two dust components.  The {\bf F}1 frequency configuration only covers
frequency $\ga 90 $GHz and therefore has the poorest capability of separating the
synchrotron component from others. Therefore should have the worst
performance. The ABS output is unbiased only for very low instrument noise ($\sigma_{\mathcal{D}}^{\rm
  inst}\la 3\times 10^{-5}\muk^2$). Systematic bias quickly grows with
increasing instrument noise. When
$\sigma_{\mathcal{D}}^{\rm inst}=0.01 \mathcal{D}_{\rm B}=5\times
10^{-5}\muk^2$,  the bias is already $20\%$ and the significance is
$1\sigma$ (foreground model C). The bias quickly increases to $40\%$ when
$\sigma_{\mathcal{D}}^{\rm inst}=10^{-3}\muk^2$, and becomes 
statistically significant ($2.5\sigma$).  The fractional bias remains
roughly a constant for larger $\sigma_{\mathcal{D}}^{\rm inst}$, but
its significance becomes weaker due to increasing statistical
error. 

This bias decreases with decreasing synchrotron amplitude, but it can
still be statistically significant even when the synchrotron is only $10\%$ of the observational upper
limit. It is therefore  dangerous to neglect the possible synchrotron
foreground.  Adding more frequency channels can significantly improve the
situation (Fig. \ref{fig:S1234}).  Adding a $35$ GHz band (the {\bf
  F}2 configuration), the bias vanishes when $\sigma^{\rm
  inst}_{\mathcal{D}}\la 3\times 10^{-4}\mu{\rm K}^2$. Further adding a $353$
band (the {\bf F}3 configuration), the bias completely
disappears and the performance of ABS is similar to the {\bf F}0
frequency configuration.

\subsection{Survey requirements}
The above tests show the following behaviors about the observed. (1)
Incomplete frequency coverage alone may not necessarily lead to biased
B-mode measurement, unless the instrument noise exceeds certain threshold
$\sigma_{\mathcal{D}}^{\rm thres}$. (2)  The bias is  positive,
increases with $\sigma_{\mathcal{D}}^{\rm inst}$ until 
reaching a plateau. These behaviors, along with the values of
$\sigma_{\mathcal{D}}^{\rm thres}$ and the bias, can be well understood within the framework of the ABS
method. They actually reflect the limitation of a given CMB
experiment.


\bfi{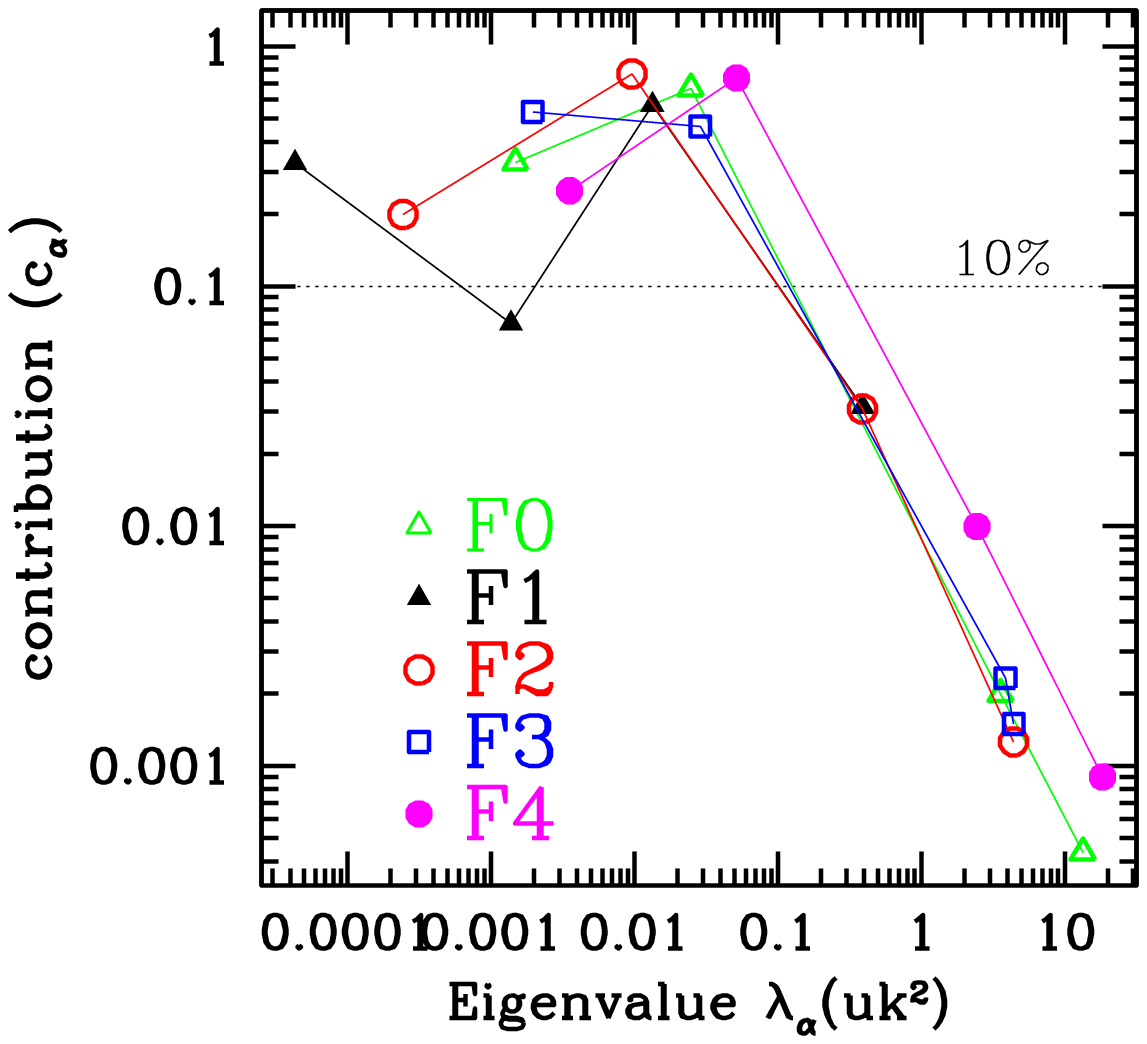}
\caption{A diagnostic of unbiased B-mode
  extraction. The detection significance of the $\alpha$-th eigenmode
  is  $\lambda_\alpha/\sigma^{\rm inst}_{\mathcal{D}}$. If $\lambda_\alpha\la
  \sigma^{\rm inst}_{\mathcal{D}}$, this eigenmode becomes
  in-detectable. This results in a fractional systematic error of $c_\alpha/(1-c_\alpha)$ in 
  $\mathcal{D}_{\rm B}$. Incomplete frequency coverage leads to the
  existence of physical eigenmodes with small $\lambda_\alpha$, but
  significant $c_\alpha$. \label{fig:ev}}
\efi
We define 
\ba
\label{eqn:contribution}
c_\alpha \equiv \frac{G_\alpha^2/\lambda_\alpha}{\sum_{\mu=1}^{M+1}
  G_\mu^2/\lambda_\mu}\ .
\ea
This is essentially the contribution of the $\alpha$-th eigenmode to
the measurement of $\mathcal{D}_{\rm B}+\mathcal{S}$
(Eq. \ref{eqn:analyticalsolution} \& \ref{eqn:shift}). If we miss
this eigenmode, the ABS determined $\mathcal{D}_{\rm B}$ will be
biased up by
\ba 
\label{eqn:bias}
\frac{\delta \mathcal{D}_{\rm B}}{\mathcal{D}_{\rm B}}=
\frac{c_\alpha}{1-c_\alpha}\times
\left(1+\frac{\mathcal{S}}{\mathcal{D}_{\rm B}}\right)\equiv b_\alpha>0\ .
\ea

The necessary condition of unbiased $\mathcal{D}_{\rm B}$  measurement by
a given survey is that all eigenmodes of significant $b_\alpha$ must be robustly 
identified. The S/N of the $\alpha$-th eigenmode is
$\tilde{\lambda}_{\alpha}$.   It is also the detection significance of this
eigenmode.  For our simplified case with identical
instrument noise level across frequency bands, $\tilde{\lambda}_\alpha=\lambda_\alpha/\sigma^{\rm
  inst}_{\mathcal{D}}$. Therefore, if $\lambda_\alpha\la \sigma^{\rm
  inst}_{\mathcal{D}}$, this eigenmode is overwhelmed by instrument
noise and in-detectable. Inappropriate frequency coverage leads to the existence of such
eigenmode with significant $b_\alpha$ but tiny $\lambda_\alpha$. It
then causes significant overestimation of
$\mathcal{D}_{\rm B}$.

Fig. \ref{fig:ev} shows $\lambda_\alpha$-$c_\alpha$ in the {\bf
  F}0-{\bf F}4 configurations for foreground case C. The CMB signal is
$\mathcal{D}_{\rm B}=5\times 10^{-3}\mu{\rm K}^2$. The shift parameter
$\mathcal{S}=0$ so $c_\alpha=b_\alpha$. Fig. \ref{fig:shift}
shows the dependence of  $\lambda_\alpha$ and $c_\alpha$  on the shift
parameter $\mathcal{S}$. $\mathcal{D}_{ij}$
of case C has $M+1=4$ physical eigenmodes. The first two  are
usually dominated by foregrounds and therefore have large eigenvalues. But due to the $1/\lambda_\mu$
weighting in Eq. \ref{eqn:analyticalsolution}, their impacts on
the B-mode extraction are automatically suppressed to a level
negligible ($b_\alpha\ll 1$) .  Usually both the third and fourth eigenmodes have significant
$c_\alpha$, and therefore are important for B-mode extraction. 
The  problem of {\bf F}1 is that the fourth eigenmode has a large $c_4=0.33$ but a
tiny $\lambda_4=4.3\times 10^{-5}\mu{\rm K}^2$.  The operation of
Eq. \ref{eqn:DB1} with $\mathcal{S}>0$ changes this eigenvalue, but $b_4$ is
essentially unchanged (Fig. \ref{fig:shift}).  Missing
this eigenmode then biases $\mathcal{D}_{\rm B}$ up by $b_4\simeq 
50\%$. This explains the observed bias for the ${\bf F}$1 frequency
configuration, when $\sigma^{\rm inst}_{\mathcal{D}}\ga \lambda_4$
(Fig. \ref{fig:S1234}).

Nevertheless, $\mathcal{S}>0$ benefits the determination of CMB B-mode. The eigenvalue increases with
increasing $\mathcal{S}$ (Fig. \ref{fig:shift}). Therefore this eigenmode
becomes more significant against instrument noise. For fixed
instrument noise, it leads to reduced systematic error. 

Adding the 35 GHz band ({\bf F}2) improves the
identification of synchrotron foreground. It leads to significantly larger 
$\lambda_4=2.4\times 10^{-4}\mu{\rm K}^2$ and significantly smaller $c_4=0.2$. This
significantly improves the situation, until $\sigma^{\rm inst}_{\mathcal{D}}\ga
\lambda_4$, where a bias of $b_4\simeq 25\%$ develops. An extra $353$ GHz band
pushes all $\lambda_\alpha>10^{-3}\mu{\rm k}^2$ and the systematic
error essentially vanishes.    

\bfi{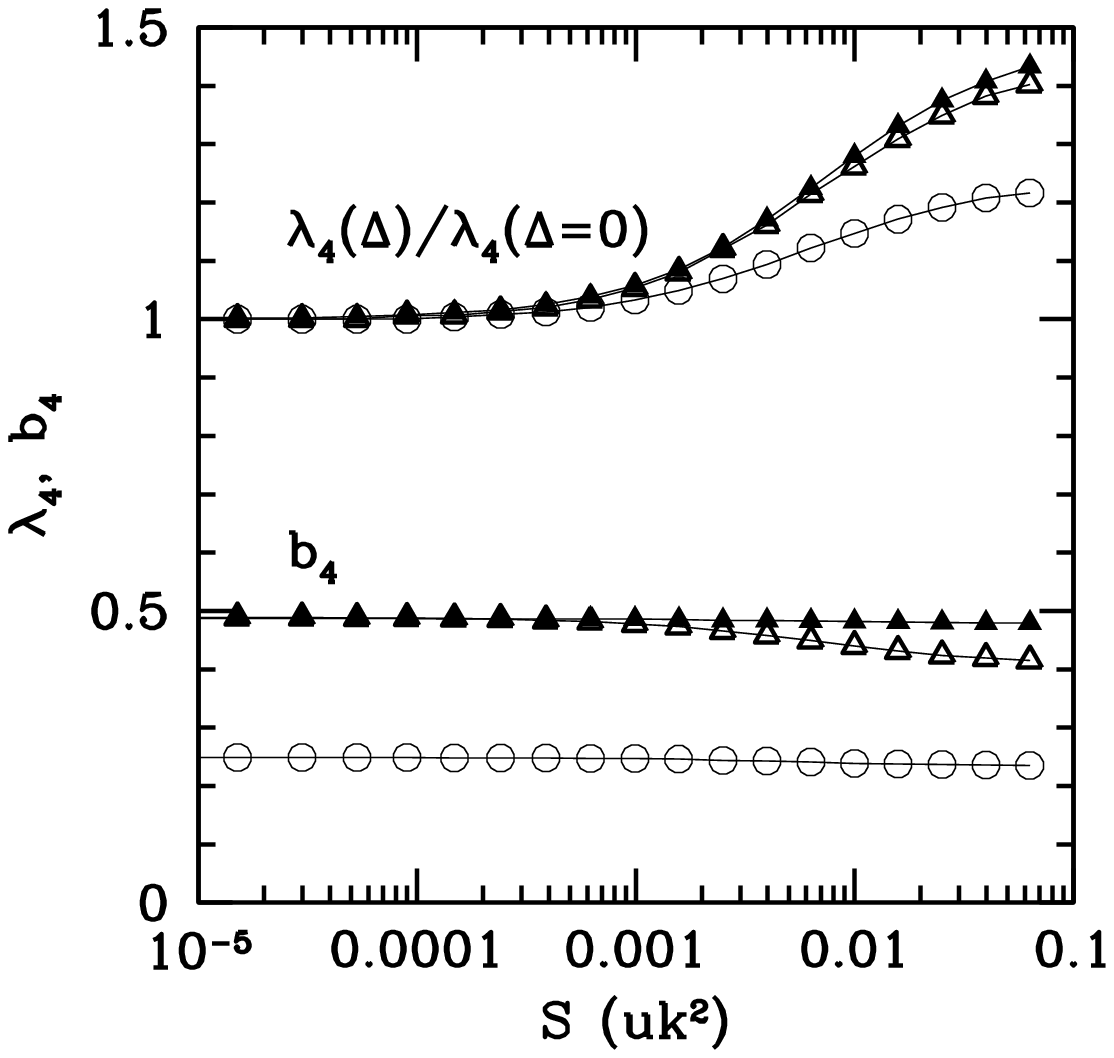}
\caption{The dependence of the smallest eigenmodes on the shift
  parameter $\mathcal{S}$, for foreground model C and frequency
  configuration ${\bf F}0$-$3$ . \label{fig:shift}}
\efi

Adding more  frequency bands further reduces the risk of more complicated
foregrounds.  Future experiments such as CORE, PRISM and PIXIE 
will have  dozens or more  frequency bands, and low instrument noise
($\sigma_{\mathcal{D}}^{\rm inst}\la 10^{-4}\mu {\rm K}^2$). Such high
degree of redundancy would make them safe for even more complicated
foregrounds. We expect them to achieve unbiased and precise
measurement of $\mathcal{D}_{\rm B}$ (e.g. {\bf F}4,
Fig. \ref{fig:S1234}).   

Therefore Eq. \ref{eqn:contribution} \& \ref{eqn:bias} can be useful
for survey design. Given the fiducial foreground and survey
configurations, these equations tell us whether this survey is
sufficient for B-mode detection, what survey depth is required, and
what  the gain of adding extra frequency channels is. 

\section{Discussions and Conclusions}
\label{sec:conclusion}
We demonstrated that the ABS method passes various tests described in the
paper. We then expect that it is applicable to general case of
foregrounds for blind, yet accurate,  extraction of CMB. It is
also numerically stable against various instrument  noise.  It further provides a
quantitative requirement (Eq. \ref{eqn:contribution}) for unbiased
B-mode measurement, useful for design of future
CMB experiments. 
 Nevertheless, we address a few caveats of these tests. (1)
These tests are performed under simplified conditions. For  example,
it neglects survey maks and beams. In the appendix, we 
discuss that the ABS method also applies when survey windows/masks
exist. (2) The simulated data used for our tests includes both
synchrotron and thermal dust foreground and takes 
decorrelation of dust foreground into account. However, it does not
include other possible foregrounds such as spinning 
dust, due to large uncertainty in their understandings. Tests against simulated
Planck temperature maps have demonstrated its robustness against more  foregrounds.
\citep{2018arXiv180707016Y}. In future
works we will extend this work to polarization maps, include more foregrounds, consider more realistic
(and therefore more complicated) instrumental noise, and redo the tests
carried out in this paper.    (3) With the presence of instrument
  noise, the ABS method contains two parameters $\lambda_{\rm
    cut}$ and $\mathcal{S}$, which are not completely fixed. The first is to exclude unphysical eigenmodes
  generated by instrument noise. The second is to reduce the impact of
instrument noise to the CMB signal. Although we are not able to fix them
unambiguously, we have physically motivated argument for $\lambda_{\rm
cut}\sim 1$. We also have a procedure to determine $\mathcal{S}\sim 10
\sigma_{\mathcal{D}}^{\rm inst}$ from data alone.  These choices
indeed work, as demonstrated by various tests. Nevertheless,  We are
still lack of more quantitative method to fix the optimal choices. The
role of $\lambda_{\rm cut}$ is to identify physical eigenmodes,
therefore the Akaike information criterion used in GNLIC (generalized
needlet ILC \citep{2011MNRAS.418..467R, 2016arXiv160509387P}) to
determine the number of physical eigenmodes may
provide an alternative.\footnote{We thank an anonymous referee for
  this point. } The Bayesian information criterion may be
explored as well.  We leave these important issues for future investigation.

The ABS method shares some similarities with the  existing method
SMICA and ILC,  while has its own uniqueness.  SMICA (spectral matching
ICA) is a version of ICA \citep{2003MNRAS.346.1089D,SMICA}). Both SMICA and ABS work directly
at the level of power spectrum, and solve essentially the same
equations (Eq. \ref{eqn:Dij}). SMICA 
simultaneously fits many unknown parameters of CMB and foregrounds against the power spectrum
measurements of all frequency bands and multipole bins.  It has been 
successfully applied to the Planck data
(e.g. \citet{2015arXiv150205956P}). SMICA has the advantage of
simultaneously solving for both CMB and foregrounds. It also has the
advantage of fitting the instrument noise power spectrum, in case that
one can not accurately calibrate/know the instrument noise from TOD
data.   ABS has the sole goal 
of solving for the CMB power spectrum, and therefore itself provides
no information on foregrounds. The advantage is that it is based on
the discovered analytical  solution
(Eq. \ref{eqn:analyticalsolution}), and computationally straightforward.  Therefore the two methods
are highly complementary. For example, the two solutions of CMB
solved by SMICA and ABS provide natural cross-checks to each
other. The CMB solution provided by ABS can be used as CMB prior in SMICA to alleviate computational
challenges in multiple parameter fitting.  On the other hand, SMICA
can identify foregrounds and  provides useful information on the
applicability of ABS, which requires $M<N_f$ to be unbiased. SMICA
also provides a check whether we understand the instrument noise
correctly and therefore  if we subtract the noise ensemble average
correctly in the ABS method.  

The ILC method and various versions of it
(e.g. \citet{2003ApJS..148...97B, Tegmark03,2004ApJ...612..633E,2011MNRAS.418..467R,2012MNRAS.419.1163B,2013MNRAS.435...18B})
were originally designed to minimize the variance in the 
reconstructed CMB map. Since foregrounds, in contrast to instrument noise, are fixed realizations of
random processes, the reconstructed map is usually biased by residual
foregrounds
\citep{2004ApJ...612..633E,2008A&A...487..775V}. \citet{2008A&A...487..775V, 2008PhRvD..78b3003S}
proved that, when the number $M$ of independent foregrounds (at map level)
is smaller than the number $N_f$ of frequency bands, and when instrument
noise is negligible, the reconstructed map is unbiased. 
On the other hand, the ABS method is designed to achieve unbiased
CMB power spectrum measurement, instead of minimizing variance in the
reconstructed CMB map.  The
analytical solution (Eq. \ref{eqn:analyticalsolution}) is derived
directly from this specific goal.  Under the condition of  no instrument noise and $M< N_f$,  the ABS method
is equivalent to the ILC method in harmonic space \citep{Tegmark03,
  2008A&A...487..775V, 2008PhRvD..78b3003S}). In this case, ABS
provides a proof of unbiased CMB reconstruction by ILC, independent of
proofs in the literature \citep{2008A&A...487..775V,
  2008PhRvD..78b3003S}.  

In reality,  instrument noise exists.  The difference between ABS and ILC
increases with  the noise-to-signal ratio of CMB experiment. It is significant
for B-mode measurement, since for relevant CMB experiments the
instrument noise is at least comparable to the elusive primordial
B-mode signal. (1) The CMB power spectrum directly obtained from
the ILC reconstructed map is $\mathcal{D}^{\rm ILC}_{\rm B}=1/({\bf
  f}_{\rm B}({\bf
  \mathcal{D}^{\rm obs}}+{\bf \langle N\rangle})^{-1}{\bf f}_{\rm B}^T)$. Here
$\mathcal{D}^{\rm obs}$ is defined in Eq. \ref{eqn:Dobs}. ${\bf \langle
  N\rangle}$ is
the ensemble average of the noise matrix, which is full
rank. This estimate of CMB power spectrum is biased and some
de-biasing procedures are required to obtain unbiased power spectrum measurement \citep{2008PhRvD..78b3003S,2012arXiv1203.4837D}. As a comparison,  the CMB power spectrum obtained by the ABS
method with the shift parameter $\mathcal{S}=0$ is $\mathcal{D}^{\rm ABS}_{\rm B}=1/({\bf
  f}_{\rm B}({\bf  
  \mathcal{D}^{\rm obs}})_{\rm pseudo}^{-1}{\bf f}_{\rm B}^T)$.
Namely the matrix we deal with is the one subtracting the ensemble average of the noise band
power. For this reason, this matrix is not always positive definite.  There may exist negative eigenvalues when the residual instrument
noise is non-negligible comparing to at least one eigenmode of CMB
plus foreground. In contrast, the eigenvalues in the ILC method  are
always positive. A further difference is the normalization of
$\mathcal{D}^{\rm obs}$. The ABS method normalizes it by the r.m.s. of
the residual instrument noise.  Therefore eigenmodes  with eigenvalue $\lambda\la 1$
have non-negligible contamination from the instrument noise and must be
excluded.   This defines the threshold $\lambda_{\rm cut}$ in the 
pseudo-inverse of $\mathcal{D}^{\rm obs}$.  It turns out that, this cut not only
reduces systematic error, but also alleviates 
the amplification of statistical error \citep{2008A&A...487..775V}.
The ILC method has other normalizations such as the noise
covariance matrix $\langle \bf N\rangle$ (e.g. GNILC, \citet{2011MNRAS.418..467R})
or some specifically defined ``nuisance'' covariance matrix (e.g. 
\citet{2016arXiv160509387P}). With these normalizations, the eigenvalues usually have a
lower bound of unity  \citep{2011MNRAS.418..467R}. Physical eigenmodes then
have eigenvalues $1+\epsilon$ ($\epsilon>0$). $\epsilon$ determines
the number of physical eigenmodes, similiar to the role of
$\lambda_{\rm cut}$ in ABS. The exact value of
$\epsilon$ is determined case by case. \citet{2011MNRAS.418..467R}
pointed out that ambiguities in the choice of $\epsilon$ may be
alleviated by adopting the Akaike information criterion (AIC), which
has been consequently applied in the Planck analysis (Eq. 5,
\citet{2016arXiv160509387P}).  Similar procedure may be applied to ABS
as well to alleviate the ambiguity associated with the choice of
$\lambda_{\rm cut}$. (2) The  version of ABS that we recommend introduces a shift
parameter $\mathcal{S}\sim 10\sigma^{\rm inst}_{\mathcal{D}}$, which
has no analogy in ILC.  This non-zero $\mathcal{S}$ is an essential
ingredient in ABS.  In particular,  ABS with $\mathcal{S}\sim
10\sigma^{\rm inst}_{\mathcal{D}}$  is able to pass the null test. Given that the
lower bound of primordial B-mode amplitude is not constrained at all
and given that  the primary
goal of ongoing B-mode experiments is to measure this amplitude, this
null test is of crucial importance to demonstrate the robustness of
B-mode detection. We then expect that the ABS method is
complementary to existing methods such as SMICA and ILC, and provide
useful cross-checks. Nevertheless, the tests that we have carried out so
far may  still be too limited to fully explore the applicability of
the ABS method. Furthermore, we have not tested it with real data, and
therefore can not compare with sophisticated methods such as SMICA and ILC
quantitatively. We leave these investigations for future study. 

The ABS method works not only for the power spectrum reconstruction,
but also for other two point statistics such as correlation function
and  variance in pixel/wavelet space. It also works beyond the B-mode
measurements, and is applicable for blind measurements of CMB temperature, 
E-mode polarization,  the thermal SZ effect and CMB spectra
distortion. Tests against simulated Planck maps has validated its
applicability to simulated CMB temperature measurements
\citep{2018arXiv180707016Y}.  Furthermore, the ABS method has
important applications even in totally different areas. For example,
it may serve as the ultimate
solution to the original proposal of extracting cosmic magnification by 
counting galaxies \citep{Zhang05,YangXJ11,YangXJ15}. The crucial
problem that ABS solve is the stochasticity bias in the intrinsic
galaxy clustering, which is analogous to multiple CMB 
foregrounds or decorrelation within each foreground component
(e.g. thermal dust of different temperature and power index along the
same lines of sight). It is then capable of reconstructing weak lensing
to high accuracy \citep{YangXJ17,Zhang18}. Therefore we expect the ABS
method to be promising and be useful in a variety of situations, and
other potential applications  should  be explored as well. 

\section{Acknowledgments}
 We thank Xuelei Chen, Jacques Delabrouille, Simon Prunet, Xinjuan
 Yang, Yu Yu and anonymous referees for useful suggestions and
 discussions. This work was supported by the National  
Science Foundation of China (11653003,11433001,11621303,11320101002)
and National Basic Research Program of China (2015CB857001,
2013CB834900). 

\bibliographystyle{mnras}
\bibliography{B}

\appendix
\section{The derivation of the ABS method}
\subsection{{\bf Uniqueness of solution}}
We first define vectors in frequency space of $N_f$ dimensions, ${\bf
  f}^{(\alpha)}\equiv
(f^{(\alpha)}_1,\cdots,f^{(\alpha)}_{N_f})$. Without loss of
generality, we absorb $\mathcal{D}_\alpha$ into corresponding ${\bf f}^{(\alpha)}$. 
Suppose that $(\sigma, {\bf h}^{(1)},\cdots, {\bf h}^{(M)})$ is also a
set of solution to Eq. \ref{eqn:Dij} \& \ref{eqn:foreground}, 
\be
\label{eqn:h}
\mathcal{D}_{ij}=f^{\rm B}_if^{\rm B}_j\sigma+\sum_{\beta=1}^M h^{(\beta)}_ih^{(\beta)}_j\ .
\ee
Then, ${\bf h}^{(\beta)}$ must be linear combinations of the
eigenvector ${\bf E}$s. Since ${\bf E}$s are linear combinations of
vector ${\bf f}^{(\alpha)}$ and ${\bf f}^{\rm B}$, ${\bf h}^{(\beta)}$ must be
linear combinations of ${\bf f}^{(\alpha)}$ and ${\bf f}^{\rm B}$,
\be
{\bf h}^{(\beta)}=\left[\sum_{\alpha=1}^M R_{\alpha\beta}{\bf
    f}^{(\alpha)}\right]+B_\beta {\bf f}^{\rm B}\ .
\ee
Here, $R_{\alpha\beta}$ and $B_\beta$ are constants to be
determined. Plug the above relation into Eq. \ref{eqn:h} and compare
with Eq. \ref{eqn:Dij} \& \ref{eqn:foreground}, we obtain
\ba
\sum_{\beta} R_{\alpha\beta}R_{\gamma\beta}=\delta_{\alpha\gamma}\
,\ \sum_\beta R_{\alpha\beta}B_\beta=0\ ,\  \mathcal{D}_{\rm B}-\sum_\beta B^2_\beta=\sigma\ . 
\ea
The first relation state that the matrix ${\bf R}$ is orthogonal,
${\bf R}^T{\bf R}=I$ where $I$ is the unity matrix. Hence det${\bf
  R}=\pm 1$. Therefore 
\be
{\rm det}{\bf R}\neq 0\  \&\ {\bf R}\cdot {\bf B}=0
\Rightarrow {\bf B}=0\Rightarrow \sigma=\mathcal{D}_{\rm B}\ .
\ee
We then prove that  the solution to $\mathcal{D}_{\rm B}$ is unique. 

In contrast, solutions to ${\bf f}^{(\alpha)}$ are not unique,
subject to transformation defined by ${\bf R}$ with det ${\bf
  R}=\pm 1$. Actually when det ${\bf R}=1$, ${\bf R}$ is the unitary rotation
matrix operating in the $M$ dimension frequency space.  It is only
after we fix the physics of each  ${\bf f}$s, may we uniquely solve
them. 

\subsection{{\bf Analytical Solution of $\mathcal{D}_{\rm B}$}}
\label{sec:analyticalsolution}
From  Eq. \ref{eqn:Dij}, we obtain $E_{\mu\nu}=G_\mu G_\nu
\mathcal{D}_{\rm B} +F_{\mu\nu}$. 
Here, $F_{\mu\nu}\equiv \sum_{ij} E^{(\mu)}_i
\mathcal{D}^{\rm fore}_{ij}E^{(\nu)}_j $ and $ G_\mu\equiv \sum_i f^{\rm B}_i E^{(\mu)}_i$.  
$E_{\mu\nu}$ is diagonal ($E_{\mu\nu}=\lambda_\mu\delta_{\mu\nu}$),
with order $M+1$ and rank $M+1$. Moving $G_\mu G_\nu
\mathcal{D}_{\rm B}$ to the l.h.s., we obtain
\be
E_{\mu\nu}-G_\mu G_\nu \mathcal{D}_{\rm B}=F_{\mu\nu}\ .
\ee
The rank of {\bf F} is $M$, smaller than its order $M+1$. As a
result, 
\be
\label{eqn:det}
{\rm det} \left({\bf E}-{\bf G}{\bf G}^T\mathcal{D}_{\rm
    B}\right)=0\ .
\ee
Here ${\bf G}$ is a column vector. 
The Sylvester's determinant theorem states that for matrices A ($m\times n$),
B($n\times m$), X($m\times m$) and unitary matrix $I_n$ ($n\times n$),
\be 
{\rm det}({\bf X}+{\bf A}{\bf B})={\rm det}({\bf X}){\rm
  det}({\bf I}_n+{\bf B}{\bf X}^{-1}{\bf A}) \ .
\ee 
Eq. \ref{eqn:det} then becomes
\ba
\label{eqn:Sylvester}
0={\rm det}\left({\bf E}-{\bf G}{\bf G}^T\mathcal{D}_{\rm
    B}\right)={\rm
det}({\bf E})\left({\bf I}_1-\mathcal{D}_{\rm B}{\bf G}^T{\bf E}^{-1}{\bf G}\right) \ . \no
\ea
Since det$({\bf E})\neq 0$, we prove Eq. \ref{eqn:analyticalsolution}. It
also proves the uniqueness of the solution for $\mathcal{D}_{\rm B}$
from Eq. \ref{eqn:Dij}.  We emphasize that the above proof is obtained under
the condition that the CMB vector is not contained in the subspace
extended by the $M$ foreground  eigenvectors. This is the prerequisite
of extracting CMB from foregrounds with only frequency information. 

\subsection{{\bf The error estimation}}
\label{sec:error}
The measured band powers are subject to instrumental noise. After
subtracting the ensemble average from the diagonal elements, there
will still be random noise $\delta \mathcal{D}^{\rm inst}_{ij}$ (with 
$\langle \delta \mathcal{D}^{\rm inst}_{ij}\rangle=0)$ on top of
$\mathcal{D}_{ij}$ of Eq. \ref{eqn:Dij}.  
In the limit of small perturbations,
\ba
\delta \lambda_\mu=\delta \mathcal{D}_{\mu\mu}\ , \ 
\delta {\bf E}^{(\mu)}= \sum_{\nu\neq \mu} \frac{\delta
  \mathcal{D}_{\mu\nu}}{\lambda_\mu-\lambda_\nu} {\bf E}^{(\nu)}\ . 
\ea
Here $\delta \mathcal{D}_{\mu\nu}\equiv \sum_{ij}
E^{(\mu)}_i \delta\mathcal{D}^{\rm inst}_{ij} E^{(\nu)}_j$. Correspondingly, 
\be
\delta G_\mu=\sum_{\nu\neq \mu} \frac{\delta
  \mathcal{D}_{\mu\nu}}{\lambda_\mu-\lambda_\nu} G_{\nu}\ ,\ 
\delta E_{\mu\nu}=\delta \lambda_\mu \delta_{\mu\nu}\ . \no
\ee
Here we have required the eigenvectors to be normalized to unity, and
for that,  $E_{\mu\nu}=\lambda_\mu\delta_{\mu\nu}$.  We
obtain by perturbing Eq. \ref{eqn:analyticalsolution}, 
\ba
\frac{\delta \mathcal{D}_{\rm B}}{\mathcal{D}_{\rm B}^2}&=&-\sum_\mu (2\delta G_\mu
\lambda_\mu^{-1}G_\mu-\lambda_\mu^{-2}\delta\lambda_\mu G_\mu^2) \no
\\
&=&-\sum_\mu \sum_{\nu\neq \mu} \frac{2\delta \mathcal{D}_{\mu\nu}G_\mu
  G_\nu}{(\lambda_\mu-\lambda_\nu)\lambda_\mu}+\sum_\mu
\lambda_\mu^{-2}\delta \mathcal{D}_{\mu\mu} G_\mu^2\ . 
\ea
We now symmetrize the first term on the right side of the above
equation. We can switch between $\mu\leftrightarrow \nu$ and take the average,
\ba
\frac{1}{2}\left(-\sum_\mu \sum_{\nu\neq \mu} \frac{2\delta \mathcal{D}_{\mu\nu}G_\mu
  G_\nu}{(\lambda_\mu-\lambda_\nu)\lambda_\mu} -\sum_\nu \sum_{\mu\neq \nu} \frac{2\delta \mathcal{D}_{\mu\nu}G_\mu
  G_\nu}{(\lambda_\nu-\lambda_\mu)\lambda_\nu}\right) \no \\
=\sum_{\mu,\nu\neq \mu} \frac{\delta \mathcal{D}_{\mu\nu}G_\mu
  G_\nu}{\lambda_\mu \lambda_\nu}\ . 
\ea
Therefore
\be
\frac{\delta \mathcal{D}_{\rm B}}{\mathcal{D}_{\rm B}^2}=\sum_{\mu,\nu} \frac{\delta \mathcal{D}_{\mu\nu}G_\mu
  G_\nu}{\lambda_\mu \lambda_\nu}\ . 
\ee
Notice that now the sum includes $\mu=\nu$
pairs. $\sigma^2_\mathcal{D}\equiv \langle \delta \mathcal{D}_{\rm
  B}^2\rangle$ can be derived  using the following relation, 
\ba
\langle \delta \mathcal{D}_{\mu\nu}\delta \mathcal{D}_{\rho\sigma}\rangle=\sum_{ijkm}
E^{(\mu)}_iE^{(\nu)}_j  E^{(\rho)}_kE^{(\sigma)}_m\langle \delta
\mathcal{D}_{ij}\delta \mathcal{D}_{km}\rangle \\
=\frac{1}{2}\sum_{ij}
E^{(\mu)}_iE^{(\nu)}_j\sigma^{\rm inst}_{\mathcal{D},i}\sigma^{\rm inst}_{\mathcal{D},j}
(E^{(\rho)}_iE^{(\sigma)}_j+E^{(\sigma)}_iE^{(\rho)}_j) \no\ .
\ea 
The last expression uses the relation $\langle \delta \mathcal{D}_{ij}\delta
\mathcal{D}_{km}\rangle=\sigma^{\rm
  inst}_{\mathcal{D},i}\sigma^{\rm
  inst}_{\mathcal{D},j}(\delta_{ik}\delta_{jm}+\delta_{im}\delta_{jk})/2$. 
The prefactor $1/2$ ensures that $\langle (\delta
\mathcal{D}_{ii})^2\rangle=(\sigma^{\rm inst}_{\mathcal{D},i})^2$. For
the simplified case $\sigma^{\rm inst}_{\mathcal{D},1}=\sigma^{\rm
  inst}_{\mathcal{D},2}=\cdots=\sigma^{\rm inst}_{\mathcal{D}}$, we
have
\ba
\sigma_{\mathcal{D}}=\sigma^{\rm inst}_{\mathcal{D}} \left(\sum_{\mu}
\frac{G_\mu^2}{\lambda_\mu^2} \mathcal{D}_{\rm B}^2\right)\ .
\ea
For general case of $\sigma^{\rm inst}_{\mathcal{D}}$ varying with frequency, it is 
\ba
\sigma_{\mathcal{D}}=\sum_{\mu}
\frac{\tilde{G}_\mu^2}{\tilde{\lambda}_\mu^2}\mathcal{D}_{\rm B}^2\ .
\ea
The above two results are for the case of $\mathcal{S}=0$. When
$\mathcal{S}\neq 0$, the factor $\mathcal{D}_{\rm B}$ shall be
replaced by $\mathcal{D}_{\rm B}+\mathcal{S}$. 

\section{The applicability of ABS in real surveys}
\label{sec:appendixB}

The ABS method has been derived under simplified situations. So an
immediate question is whether it can be applied to real CMB
surveys. The main text incorporate the fact that different frequency
bands have different instrument noises and  
therefore should have different weights to obtain optimal measurement
of CMB B-mode.  There are other complexities.  The measured CMB is smoothed over the beam, which
depends on frequency. The masks in general vary with frequency. Even
if we adopt identical mask for all frequency bands, the interplay
between beam and mask causes decorrelation of the CMB signal in different frequency
bands. Here we outline a procedure to apply the ABS method with the
presence of these complexities.  
\bi
\item {\bf Step 1}. We smooth all maps to a fiducial beam $B^f(\theta)$,
  before masking. This beam should be identical for all frequency
  bands. It should be homogeneous and isotropic, for
  the convenience of later process. Therefore it should only depend on
  the angle $\theta$ between the pixel position ($\hat{n}_{\rm
   pixel}$), and the sky position ($\hat{n}$) where the signal comes from. If the actual beams are
  also homogeneous,  this step can be done efficiently in harmonic space by
  multiplying $a_{lm}$ of the $i$-th frequency band by
  $B^f(l)/B_i(l,m)$. Here $B_i(l,m)$ is the beam of the $i$-th
  frequency band, which can be anisotropic. In reality, the beam is in general inhomogeneous
  (depending on $\hat{n}_{\rm pixel}$) and this step shall be done
  pixel by pixel. 
\item {\bf Step 2}. We chose and apply a common mask $M(\hat{n}_{\rm
    pixel})$ for all frequency maps. 
\item {\bf Step 3}. We then measure the cross band power between these
  smoothed/masked maps $\mathcal{D}^{{\rm S}+{\rm M}}_{ij}(\ell)$. We
  subtract the ensemble average of the instrument noise power spectra from the diagonal
  elements. We also estimate the r.m.s. of the residual instrument
  noise  $\sigma_{\mathcal{D},i}^{\rm inst}(\ell)$ in each map. 
\item {\bf Step 4}.  We then weigh
  $\mathcal{D}^{{\rm S}+{\rm M}}_{ij}$ by $\sqrt{\sigma^{\rm
      inst}_{\mathcal{D},i}\sigma^{\rm inst}_{\mathcal{D},j}}$ and obtain
  essentially the S/N matrix $\tilde{\mathcal{D}}_{ij}$. The ABS method then directly applies to
\be
\label{eqn:absmodified}
\tilde{\mathcal{D}}_{ij}\equiv
\frac{\mathcal{D}^{{\rm S}+{\rm M}}_{ij}}{\sqrt{\sigma^{\rm
      inst}_{\mathcal{D},i}\sigma^{\rm inst}_{\mathcal{D},j}}}=\tilde{f}^B_i\tilde{f}^B_j\mathcal{D}^{{\rm S}+{\rm M}}_{\rm
  B}+\tilde{\mathcal{D}}^{\rm fore}_{ij}\ .
\ee
Here, $\tilde{f}^{\rm B}_i\equiv f^{\rm B}_i/\sqrt{\sigma^{\rm
    inst}_{\mathcal{D},i}}$. $\mathcal{D}^{{\rm S}+{\rm M}}_B$ is the
CMB band power, with beam $B^f(\theta)$ and mask $M$.  The ABS method directly applies to the
above equation and solves for $\mathcal{D}^{{\rm S}+{\rm M}}_B$.
\item {\bf Step 5}.  It is then the standard procedure to deconvolve
  $\mathcal{D}^{{\rm S}+{\rm M}}_{\rm B}$ for
$\mathcal{D}_{\rm B}$ (e.g. \citet{2002ApJ...567....2H}). 
\ei
In particular, step 1 (smoothing) and step 2 (masking) are not
interchangeable. Otherwise the
CMB signal in different maps will not be linearly proportional to each
other and the signal term in Eq. \ref{eqn:absmodified} does not have
the form $\propto \tilde{f}^{\rm B}_i\tilde{f}^{\rm B}_j$.

\end{document}